\documentstyle[epsfig,aps]{revtex}

\begin{document}
\draft
\title{Polarization of the Fulling-Rindler vacuum by uniformly accelerated mirror}
\author{Aram A. Saharian\footnote{%
E-mail: saharyan@server.physdep.r.am}}
\address{Department of Physics, Yerevan State University,
1 Alex Manoogian St., 375049 Yerevan, Armenia}
\date{\today}
\maketitle
\begin{abstract}
Positive frequency Wightman function and vacuum expectation values
of the energy-momentum tensor are computed  for a massive scalar
field with general curvature coupling parameter and satisfying
Robin boundary condition on a uniformly accelerated infinite
plate. The both regions of the right Rindler wedge, (i) on the
right (RR region) and (ii) on the left (RL region) of the plate
are investigated. For the case (ii) the electromagnetic field is
considered as well. The mode summation method is used with
combination of a variant of the generalized Abel-Plana formula.
This allows to present the expectation values in the form of a sum
of the purely Rindler and boundary parts. Near the plate surface
the vacuum energy-momentum tensor is dominated by the boundary
term. At large distances from the plate and near the Rindler
horizon the main contribution comes from the purely Rindler part.
In the RL region the vacuum energy density of the electromagnetic
field is negative near the horizon and is positive near the plate.

\end{abstract}


\section{Introduction} \label{sec:introd}

The influence of boundaries on the vacuum state of a quantum field
leads to interesting physical consequences. Well known example is
the Casimir effect (see \cite{Mostepanenko,Plunien,Milton} and
references therein), when the modification of the zero-point
fluctuations spectrum by the presence of boundaries induces vacuum
forces acting on these boundaries. Another interesting effect is
the creation of particles from the vacuum by moving boundaries
\cite{Birrel}. In this paper we will study the scalar vacuum
polarization brought about by the presence of infinite plane
boundary moving by uniform acceleration through the
Fulling-Rindler vacuum. This problem for the conformally coupled
Dirichlet and Neumann massless scalar and electromagnetic fields
in four dimensional spacetime was considered by Candelas and
Deutsch \cite{CandD}. Here we will investigate the vacuum
expectation values of the energy-momentum tensor for the massive
scalar field with general curvature coupling and satisfying Robin
boundary condition on the infinite plane in an arbitrary number of
spacetime dimensions. The dimensional dependence of physical
quantities is of considerable interest in the Casimir effect and
is investigated for various geometries of boundaries and boundary
conditions (see
\cite{Amb83},\cite{Milton},\cite{Mil94},\cite{RomSah},\cite{Sahsph},\cite{Sahcyl}
and references therein). The Robin boundary conditions are an
extension of the ones imposed on perfectly conducting boundaries
and may, in some geometries, be useful for depicting the finite
penetration of the field into the boundary with the "skin-depth"
parameter related to the Robin coefficient \cite{Mostep}. It is
interesting to note that the quantum scalar field satisfying Robin
condition on the boundary of a cavity violates the Bekenstein's
entropy-to-energy bound near certain points in the space of the
parameter defining the boundary condition \cite{Solod}. This type
of conditions also appear in considerations of the vacuum effects
for a confined charged scalar field in external fields
\cite{Ambjorn2} and in quantum gravity
\cite{Mo},\cite{EK},\cite{Esp97}. Mixed boundary conditions
naturally arise for scalar and fermion bulk fields in the
Randall-Sundrum model \cite{Gherg}. In this model the bulk
geometry is a slice of anti-de Sitter space and the corresponding
Robin coefficient is related to the curvature scale of this space.
Unlike to Ref.\cite{CandD}, where the region of the right Rindler
wedge on the right from the barrier (in the following we will
denote this region as RR) is considered only, we investigate
vacuum energy density and stresses in both regions, including the
one between the barrier and Rindler horizon (region RL in the
following consideration). Our calculation is based on the
summation formula derived in Appendix \ref{sec:app1} on the base
of the generalized Abel-Plana formula \cite{Sahmat} (see also
\cite{Sahreview}). This allows to extract explicitly the part due
to the unbounded Rindler spacetime and to present the boundary
part in terms of strongly convergent integrals.

The paper is organized as follows. In the next section we consider
the vacuum expectation values for the energy-momentum tensor of
the scalar field in the region on the right of the uniformly
accelerated infinite plane. By using the summation formula derived
in Appendix \ref{sec:app1} these quantities are presented in the
form of a sum containing two parts. The first one corresponds to
the energy-momentum tensor for the Rindler wedge without
boundaries and is investigated in Sec. \ref{sec:Rindler}. The
second term presents the contribution brought about by the
presence of the boundary and is considered in Sec.
\ref{sec:right}. The corresponding boundary terms for the region
between the barrier and Rindler horizon are evaluated in Sec.
\ref{sec:left} for scalar and electromagnetic fields. In Appendix
\ref{sec:app2} we show that the direct evaluation of the
energy-momentum tensor in $D=1$ gives the same result as the
analytic continuation from higher dimensions. And finally, in
Appendix \ref{sec:app3}, we prove the identity used in Sec.
\ref{sec:Rindler} to see the equivalency of our expressions for
the expectation values of the energy-momentum tensor in the case
of the Fulling-Rindler vacuum without boundaries to the results
previously investigated in the literature for the conformally and
minimally coupled scalars.

\section{Vacuum expectation values and the Wightman function} \label{sec:vev}

Consider a real scalar field $\varphi (x) $ with curvature coupling
parameter $ \zeta $ in $D$-dimensional space satisfying Robin boundary
condition
\begin{equation}
(A+Bn^{i}\nabla _{i})\varphi (x)=0  \label{boundcond}
\end{equation}
on a plane boundary uniformly accelerated normal to itself. Here
$A $ and $B $ are constants, $n^i $ is the unit normal to the
plane, $\nabla _i $ is the covariant derivative operator. The
Dirichlet and Neumann boundary conditions are obtained from
(\ref{boundcond}) as special cases. Of course, all results in the
following will depend on the ratio $A/B$ only. However, to keep
the transition to the Dirichlet and Neumann cases transparent we
will use the form (\ref{boundcond}). The field equation has the
form
\begin{equation}
(\nabla _{i}\nabla ^{i}+m^{2}+\zeta R)\varphi (x)=0,  \label{fieldeq}
\end{equation}
where $R $ is the scalar curvature for the background spacetime.
The values of the curvature coupling $\zeta =0$ and $\zeta =\zeta
_{c}\equiv \frac{D-1}{4D}$ correspond to the minimal and conformal
couplings, respectively. In the case of a flat background
spacetime the corresponding metric energy-momentum tensor (EMT) is
given by (see, e.g., \cite{Birrel})
\begin{equation}
T_{ik}=(1-2\zeta )\partial _{i}\varphi \partial _{k}\varphi
+(2\zeta -1/2)g_{ik}\partial ^{l}\varphi \partial _{l}\varphi
-2\zeta \varphi \nabla _{i}\nabla _{k}\varphi +(1/2-2\zeta
)m^{2}g_{ik}\varphi ^{2}.  \label{EMT1}
\end{equation}
By using the field equation it can be seen that expression
(\ref{EMT1}) can be also presented in the form
\begin{equation}
T_{ik}=\partial _{i}\varphi \partial _{k}\varphi +\left[ \left( \zeta -\frac{%
1}{4}\right) g_{ik}\nabla _l\nabla ^l -\zeta \nabla _{i} \nabla
_{k}\right] \varphi ^{2}, \label{EMT2}
\end{equation}
and the corresponding trace is equal to
\begin{equation}
T_{i}^{i}=D(\zeta -\zeta _{c})\nabla _i\nabla ^i \varphi ^{2}+
m^{2}\varphi ^{2} .
\label{trace}
\end{equation}
By virtue of Eq. (\ref{EMT2}) for the vacuum expectation values (VEV's) of
the EMT we have
\begin{equation}
\langle 0\left| T_{ik}(x)\right| 0\rangle =\lim_{x^{\prime }
\rightarrow x}\nabla _{i}\nabla
_{k}^{\prime }G^{+}(x,x^{\prime })+\left[ \left( \zeta -\frac{1}{4}\right)
g_{ik}\nabla _{l}\nabla ^{l}-\zeta \nabla _{i}\nabla _{k}\right]
\langle 0\left| \varphi ^{2}(x)\right| 0\rangle ,  \label{vevemtW}
\end{equation}
where $|0\rangle $ is the amplitude for the vacuum state and we
introduced the positive frequency Wightman function:
\begin{equation}
G^{+}(x,x^{\prime })=\langle 0\left| \varphi (x)\varphi (x^{\prime })
\right| 0\rangle .
\label{Wight1}
\end{equation}
In Eq. (\ref{vevemtW}) instead of this function one can choose any
other bilinear function of fields such as the Hadamard function,
Feynman's Green function, etc. The regularized vacuum EMT does not
depend on the specific choice. The reason for our choice of the
Wightman function is that this function also determines the
response of the particle detectors in a given state of motion. By
expanding the field operator over eigenfunctions and using the
commutation relations one can see that
\begin{equation}
G^{+}(x,x^{\prime })=\sum_{\alpha }\varphi _{\alpha }(x)\varphi _{\alpha
}^{\ast }(x^{\prime }),  \label{Wightvev}
\end{equation}
where $\alpha $ denotes a set of quantum numbers, $\left\{ \varphi _{\alpha
}(x)\right\} $ is a complete set of solutions to the field equation (\ref
{fieldeq}) satisfying boundary condition (\ref{boundRind}).

Below we will assume that the plate is situated in the right
Rindler wedge, $x_{1}>\left| t\right| $. It is convenient to
introduce Rindler coordinates $(\tau ,\xi ,{\mathbf x})$ related
with Minkowski coordinates $(t,x_1,{\mathbf x})$ by
\begin{equation}
t=\xi \sinh \tau ,\quad x_{1}=\xi \cosh \tau ,  \label{Rindtrans}
\end{equation}
where ${\mathbf x}=(x_2,...,x_D)$ denotes the set of coordinates parallel
to the plate. In these coordinates a wordline defined by $\xi ,{\mathbf x}=
{\rm const}$ describes an observer with constant proper acceleration
$\xi ^{-1}$. The Minkowski line element restricted to the Rindler
wedge is
\begin{equation}
ds^{2}=\xi ^{2}d\tau ^{2}-d\xi ^{2}-d\mathbf{x}^{2}.  \label{interval}
\end{equation}
The corresponding metric is static, admitting the Killing vector
field $\partial /\partial \tau $. The Fulling-Rindler vacuum is
the vacuum state determined by choosing positive frequency modes
to have positive frequency with respect to this Killing vector.

First we will consider the region on the right from the boundary,
$\xi \geq a$, RR region, where $a^{-1}$ is the proper acceleration
of the barrier. In Rindler coordinates boundary condition
(\ref{boundcond}) for a uniformly accelerated mirror takes the
form
\begin{equation}
\left( A+B\frac{\partial }{\partial \xi }\right) \varphi =0,\quad \xi =a.
\label{boundRind}
\end{equation}
The problem symmetry allows to write the solutions to field
equation (\ref{fieldeq}) as
\begin{equation}
\varphi (x)=C\phi (\xi )e^{i{\mathbf{kx}}-i\omega \tau }.  \label{sol1}
\end{equation}
Substituting this into Eq. (\ref{fieldeq}) we see that the
function $\phi (\xi )$ satisfies the equation
\begin{equation}
\xi \frac{d}{d\xi }\left( \xi \frac{d\phi }{d\xi }\right) +\left( \omega
^{2}-\xi ^{2}\lambda ^{2}\right) \phi (\xi )=0,\quad \lambda =\sqrt{%
k^{2}+m^{2}}.  \label{eqxi}
\end{equation}
The corresponding linearly independent solutions are the Bessel
modified functions $I_{\pm i\omega }(\lambda \xi )$ and
$K_{i\omega }(\lambda \xi )$ with the imaginary order. It can be
seen that for any two solutions to equation (\ref{eqxi}), $\phi
^{(1)}_{\omega }(\xi )$ and $\phi ^{(2)}_{\omega }(\xi )$, the
following integration formula takes place
\begin{equation}
\int \frac{d\xi }{\xi }\phi ^{(1)}_{\omega }(\xi )
\phi ^{(1)}_{\omega '}(\xi )=\frac{\xi }{\omega ^2-\omega '^2}
\left[ \phi ^{(1)}_{\omega }(\xi )\frac{d\phi ^{(2)}_{\omega '}
(\xi )}{d\xi }-\phi ^{(2)}_{\omega '}(\xi )\frac{d\phi ^{(1)}_{\omega }
(\xi )}{d\xi }\right] . \label{intformgen}
\end{equation}
As we will see below the normalization integrals for the
eigenfunctions will have this form.

For the region $\xi >a$ a complete set of solutions that are of
positive frequency with respect to $\partial /\partial \tau $ and
bounded as $\xi \to \infty $ is
\begin{equation}
\varphi _{\alpha }(x)=CK_{i\omega }(\lambda \xi )e^{i{\mathbf{kx}}-
i\omega \tau },\quad \alpha =(\omega ,\mathbf{k}).  \label{sol2}
\end{equation}
From boundary condition (\ref{boundRind}) we find that the possible values
for $\omega $ have to be roots to the equation
\begin{equation}
AK_{i\omega }(\lambda a)+B\lambda K_{i\omega }^{\prime }(\lambda a)=0,
\label{modeeq}
\end{equation}
where the prime denotes the differentiation with respect to the
argument. This equation has infinite number of real zeros. We will
denote them by $\omega =\omega _n=\omega _n(k)$, $n=1,2,...$
arranged in ascending order: $\omega _n<\omega _{n+1}$. The
coefficient $C$ in (\ref{sol2}) is determined by the normalization
condition
\begin{equation}
\left( \varphi _{\alpha },\varphi _{\alpha ^{\prime }}\right) =
-i\int d{\mathbf x}\int \frac{d\xi }{\xi }\varphi _{\alpha }
\stackrel{\leftrightarrow }{\partial }_{\tau }\varphi ^{*}_{\alpha '}=
\delta _{\alpha \alpha ^{\prime }}  \label{normcond}
\end{equation}
with respect to the standard Klein-Gordon inner product, and the
$\xi $-integration goes over the region $(a,\infty )$. Using
integration formula (\ref{intformgen}) with $\phi _\omega
^{(1)}=\phi _\omega ^{(2)}= K_{i\omega }(\lambda \xi )$ and
Wronskian $W\{ K_{\nu }(z),I_{\nu }(z)\} =1/z$, from
(\ref{normcond}) one finds
\begin{equation}
C^{2}=\frac{\pi }{(2\pi )^D\omega _n}\left[ \int _{a}^{\infty }
\frac{d\xi }{\xi }K_{i\omega _n}^2(\lambda \xi )\right] ^{-1}=
\frac{1}{(2\pi )^{D-1}}\frac{\bar{I}_{i\omega _{n}}(\lambda a)}{\frac{%
\partial }{\partial \omega }\bar{K}_{i\omega }(\lambda a)\mid _{
\omega =\omega _{n}}},  \label{normc}
\end{equation}
where for a given function $f(z)$ we use the notation
\begin{equation}
\bar f(z)=Af(z)+bzf^{\prime }(z)=0, \quad b=B/a.  \label{barnot}
\end{equation}
From the first equality in Eq. (\ref{normc}) it follows that $C^2$
is real and positive. Now substituting the eigenfunctions
(\ref{sol2}) into (\ref{Wightvev}) for the Wightman function we
obtain
\begin{equation}
G^{+}(x,x')=
\int \frac{d{\mathbf k}}{(2\pi )^{D-1}}\,
e^{i{\mathbf k}({\mathbf x}-{\mathbf x'})}\sum_{n=1}^{\infty }\frac{%
\bar{I}_{i\omega }(\lambda a)}{\frac{\partial }{\partial \omega
}\bar{K}_{i\omega }(\lambda a)}%
K_{i\omega }(\lambda \xi)K_{i\omega }(\lambda \xi ')
e^{-i\omega (\tau -\tau ')}
\vert _{\omega =\omega _{n}}.
\label{emtdiag}
\end{equation}
For the further evolution of VEV's (\ref{emtdiag}) we can apply to
the sum over $n$ summation formula (\ref{sumformula}). As a
function $F(z)$ in this formula let us choose
\begin{equation}\label{Fztoform}
  F(z)=K_{iz}(\lambda \xi)K_{iz}(\lambda \xi ')e^{-iz(\tau -\tau
  ')}.
\end{equation}
Using the asymptotic formulas for the Bessel modified functions it
can be easily seen that condition (\ref{condf1}) is satisfied if
$a^2e^{|\tau -\tau '|}<|\xi \xi '|$. Note that this condition is
satisfied in the coincidence limit $\tau =\tau '$ for the points
in the region under consideration, $\xi ,\xi '>a$. With $F(z)$
from (\ref{Fztoform}) the contribution corresponding to the
integral term on the left of formula (\ref{sumformula}) is the
Wightman function for the Fulling-Rindler vacuum $\vert 0_R
\rangle $ without boundaries:
\begin{equation}
G^{+}_{R}(x,x')=\langle 0_{R}\vert \varphi (x)\varphi (x')\vert 0_{R}\rangle
= \frac{1}{\pi ^2}\int \frac{d{\mathbf k}}{(2\pi )^{D-1}}\,
e^{i{\mathbf k}({\mathbf x}-{\mathbf x'})}\int_{0}^{\infty }d\omega \sinh (\pi
\omega ) e^{-i\omega (\tau -\tau ')}K_{i\omega }(\lambda \xi )K_{i\omega }
(\lambda \xi ') .  \label{emtRindler}
\end{equation}
To see this note that for the right Rindler wedge the
eigenfunctions to the field equation (\ref{fieldeq}) are in form
(see, for instance, \cite{Fulling},\cite{Davies},\cite{Unruh},
\cite{CandRaine})
\begin{equation}
\varphi _{\omega \mathbf{k}}=\frac{\sqrt{\sinh \pi \omega }}{(2\pi
)^{(D-1)/2}\pi }e^{i{\mathbf{kx}}-i\omega \tau}K_{i\omega }(\lambda \xi ).
\label{Rindfunc}
\end{equation}
Substituting these modes into formula (\ref{Wightvev}) we
obtain Eq. (\ref{emtRindler}%
). Hence, taking into account this and applying summation formula
(\ref {sumformula}) to Eq. (\ref{emtdiag}) we receive
\begin{equation}
G^{+}(x,x')=G^{+}_{R}(x,x')+\langle \varphi (x)
\varphi (x')\rangle ^{(b)}, \label{WRb}
\end{equation}
where the second term on the right is induced by the existence of
the barrier:
\begin{equation}
\langle \varphi (x)\varphi (x')\rangle ^{(b)}= -\frac{1}{\pi }\int
\frac{d{\mathbf k}}{(2\pi )^{D-1}}\, e^{i{\mathbf k}({\mathbf
x}-{\mathbf x'})}\int_{0}^{\infty }d\omega \frac{\bar I_{\omega
}(\lambda a)}{\bar K_{\omega }(\lambda a)} K_{\omega }(\lambda \xi
)K_{\omega }(\lambda \xi ')\cosh [\omega (\tau - \tau ')] ,
\label{Wb}
\end{equation}
 and is finite for $\xi >a$. All divergences in the coincidence
limit are contained in the first term corresponding to the Fulling
- Rindler vacuum without boundaries. The VEV's for the EMT are
obtained substituting Eq. (\ref{WRb}) into Eq. (\ref{vevemtW}).
Similar to (\ref{WRb}) this VEV's are also in the form of the sum
containing purely Rindler and boundary parts. Firstly we will
concentrate to the first term.

\section{Vacuum EMT in the Rindler wedge without boundaries}
\label{sec:Rindler}

The Wightman function for the right Rindler wedge given by formula
(\ref {emtRindler}) is divergent in the coincidence limit
$x^{\prime }\rightarrow x $. This leads to the divergences in the
VEV for the EMT. To extract these divergences we can subtract from
Eq. (\ref{emtRindler}) the corresponding Wightman function for the
Minkowski vacuum $|0_{M}\rangle$.
By using mode sum formula (%
\ref{Wightvev}) and the standard Minkowskian eigenfunctions it easily can be
seen that the latter may be presented in the form
\begin{equation}
G_{M}^{+}(x,x^{\prime })=\int \frac{d^{D-1}\mathbf{k}}{(2\pi )^{D}}e^{i%
\mathbf{k}(\mathbf{x}-\mathbf{x}^{\prime })}K_{0}\left( \lambda \sqrt{\xi
^{2}+\xi ^{\prime 2}-2\xi \xi ^{\prime }\cosh (\tau -\tau ^{\prime })}%
\right) ,  \label{WMin}
\end{equation}
where $\xi ^{2}+\xi ^{\prime 2}-2\xi \xi ^{\prime }\cosh (\tau -\tau
^{\prime })=(x_{1}-x_{1}^{\prime })^{2}-(t-t^{\prime })^{2}$. Using the
integration formula \cite{Prudnikov2}
\begin{equation}
K_{0}\left( \lambda \sqrt{\xi ^{2}+\xi ^{\prime 2}+2\xi \xi ^{\prime }\cos b}%
\right) =\frac{2}{\pi }\int_{0}^{\infty }d\omega \,\cosh (b\omega
)K_{i\omega }(\lambda \xi )K_{i\omega }(\lambda \xi ^{\prime }),
\label{WMinint}
\end{equation}
Eq. (\ref{WMin}) can be presented in the form convenient for
subtraction:
\begin{equation}
G_{M}^{+}(x,x^{\prime })=\frac{1}{\pi ^{2}}\int \frac{d^{D-1}\mathbf{k}}{%
(2\pi )^{D-1}}e^{i\mathbf{k}(\mathbf{x}-\mathbf{x}^{\prime
})}\int_{0}^{\infty }d\omega \,\cosh \left\{ \omega \left[ \pi -i(\tau -\tau
^{\prime })\right] \right\} K_{i\omega }(\lambda \xi )K_{i\omega }(\lambda
\xi ^{\prime }).  \label{WMin1}
\end{equation}
As a result for the difference between the Rindler and Minkowskian
Wightman functions one obtains
\begin{eqnarray}
G_{R}^{+}(x,x^{\prime })-G_{M}^{+}(x,x^{\prime }) &=&-\frac{1}{\pi ^{2}}\int
\frac{d^{D-1}\mathbf{k}}{(2\pi )^{D-1}}e^{i\mathbf{k}(\mathbf{x}-\mathbf{x}%
^{\prime })}  \label{WRMdif} \\
&& \times \int_{0}^{\infty }d\omega \,e^{-\pi \omega }\cos \left[
\omega (\tau -\tau ^{\prime })\right] K_{i\omega }(\lambda \xi
)K_{i\omega }(\lambda \xi ^{\prime }).  \nonumber
\end{eqnarray}
After integrating over directions of the vector $\mathbf{k}$, in
particular, for the difference between VEV's of the field square
it follows from here
\begin{eqnarray}
\langle \varphi ^{2}(x)\rangle _{\mathrm{sub}}^{(R)} &=& \langle 0_{R}|
\varphi (x)|0_{R}\rangle -\langle 0_{M}|\varphi (x)|0_{M}\rangle
\nonumber \\
& = & \frac{-1}{2^{D-2}\pi ^{(D+3)/2}\Gamma \left( \frac{D-1}{2}\right) }%
\int_{0}^{\infty }dk\,k^{D-2}\int_{0}^{\infty }d\omega \,e^{-\pi \omega
}K_{i\omega }^{2}(\lambda \xi ).  \label{phi2dif}
\end{eqnarray}
Using formula (\ref{vevemtW}), for the corresponding difference
between the VEV's of the EMT we receive
\begin{eqnarray}
\langle T_i^k \rangle _{{\mathrm sub}}^{(R)}&=&\langle 0_R|T_i^k|0_R\rangle
-\langle 0_M|T_i^k|0_M\rangle   \label{emtRsub} \\
&=& \frac{-\delta _{i}^{k}}{2^{D-2}\pi ^{(D+3)/2}\Gamma \left( \frac{D-1}{2}%
\right) }\int_{0}^{\infty }dkk^{D-2}\lambda ^{2}\int_{0}^{\infty }d\omega
e^{-\pi \omega }f^{(i)}\left[ K_{i\omega }(\lambda \xi )\right] ,
\nonumber
\end{eqnarray}
where we have introduced the following notations
\begin{eqnarray}
f^{(0)}[g(z)] &=&\left(\frac{1}{2}-2\zeta \right)
\left| \frac{dg(z)}{dz}\right| ^{2}+%
\frac{\zeta }{z}\frac{d}{dz}\left| g(z)\right| ^2+\left[ \frac{1}{2}
-2\zeta +\left( \frac{1}{2}+2\zeta \right) \frac{\omega ^{2}}{z^{2}}
\right] \left| g(z)\right| ^2, \nonumber  \\
f^{(1)}[g(z)] &=&-\frac{1}{2}\left| \frac{dg(z)}{dz}\right| ^{2}-
\frac{\zeta }{z}%
\frac{d}{dz}\left| g(z)\right| ^2+\frac{1}{2}\left( 1-\frac{\omega ^{2}%
}{z^{2}}\right) \left| g(z)\right| ^2,  \label{fq} \\
f^{(i)}[g(z)] &=&\left( \frac{1}{2}-2\zeta \right) \left[ \left|
\frac{dg(z)}{dz}\right| ^{2}+
\left( 1-\frac{\omega ^{2}}{z^{2}}\right) \left| g(z)\right| ^2\right] -%
\frac{k^{2}}{(D-1)\lambda ^{2}}\left| g(z)\right| ^2,
\quad i=2,3,...,D,  \nonumber
\end{eqnarray}
and indices 0 and 1 correspond to the coordinates $\tau $ and $\xi
$.

We consider first the massless limit. In this limit interchanging
the orders of integration, making use of the integration formula
\cite{Prudnikov2}
\begin{equation}
{\mathcal I}_{D}(\xi )\equiv \int_{0}^{\infty }dk\,k^{D}K_{i\omega }^{2}(k\xi
)=\frac{2^{D-2}}{\xi ^{D+1}\Gamma (D+1)}\Gamma ^{2}\left( \frac{D+1}{2}%
\right) \left| \Gamma \left( \frac{D+1}{2}+i\omega \right) \right| ^{2}
\label{IntID}
\end{equation}
and the relations
\begin{eqnarray}
&& \int_{0}^{\infty }dk\,k^{D-1}K_{i\omega }(k\xi )K_{i\omega
}^{\prime }(k\xi )=-\frac{D-1}{2\xi } {\mathcal I}_{D-2}(\xi ),
\label{intID1} \\
&& \int_{0}^{\infty }dk\,k^{D}K_{i\omega }^{\prime 2}(k\xi )=
\frac{{\mathcal I}_{D-2}(\xi )}{4D\xi ^{2}}\left[ 4\omega ^{2}+
(D+1)(D-1)^2\right] , \label{intID2}
\end{eqnarray}
for the purely Rindler parts (\ref{phi2dif}) and (\ref{emtRsub})
one finds
\begin{eqnarray}
\langle \varphi ^2\rangle ^{(R)}_{{\rm sub}}&=&
-\frac{\xi ^{-D+1}}{2^{D}\pi ^{D/2+1}\Gamma (D/2)}%
\int_{0}^{\infty }d\omega \, e^{-\pi \omega }
\left| \Gamma \left( \frac{D-1}{2%
}+i\omega \right) \right| ^{2},  \label{phi2Rindm0} \\
\langle T_{i}^{k}\rangle ^{(R)}_{{\rm sub}}&=&
-\frac{\delta _{i}^{k}\xi ^{-D-1}}{2^{D}\pi ^{D/2+1}\Gamma (D/2)}%
\int_{0}^{\infty }d\omega \, \omega ^2 e^{-\pi \omega }
\left| \Gamma \left( \frac{D-1}{2%
}+i\omega \right) \right| ^{2}f_{0}^{(i)}(\omega ),
\label{emtRindm0}
\end{eqnarray}
where we have introduced notations
\begin{eqnarray}
f_{0}^{(0)}(\omega ) &=&-Df_{0}^{(1)}(\omega )=1+D(D-1)(\zeta
_{c}-\zeta )/\omega ^{2} , \label{f00}
\\
f_{0}^{(i)}(\omega ) &=&-1/D+(D-1)^{2}(\zeta _{c}-\zeta )/\omega ^{2}%
 ,\quad i=2,3,\ldots ,D.  \nonumber
\end{eqnarray}
Using the recurrence formula, the gamma function in Eqs.
(\ref{phi2Rindm0}) and (\ref{emtRindm0}) can be led to $\Gamma
(1+i\omega )$ for odd $D$ and to $\Gamma (1/2+i\omega )$ for even
$D$. The standard formulas for the latters \cite{abramowiz} allow
us to write expressions (\ref{phi2Rindm0}) and (\ref{emtRindm0})
in the form
\begin{eqnarray}
\langle \varphi ^2\rangle ^{(R)}_{{\rm sub}}& = &
-\frac{\xi ^{-D+1}}{2^{D-1}\pi ^{D/2}\Gamma (D/2)}%
\int_{0}^{\infty }\frac{\omega ^{D-2}d\omega }{e^{2\pi \omega
}+(-1)^D} \prod _{l=1}^{l_m}\left[ \left( \frac{D-1-2l}{2\omega }
\right) ^2+1 \right] , \label{phi2Rindm01} \\
\langle T_{i}^{k}\rangle ^{(R)}_{{\rm sub}}& = &
-\frac{\delta _{i}^{k}\xi ^{-D-1}}{2^{D-1}\pi ^{D/2}\Gamma (D/2)}%
\int_{0}^{\infty }\frac{\omega ^Dd\omega }{e^{2\pi \omega
}+(-1)^D} f_{0}^{(i)}(\omega )\prod _{l=1}^{l_m}\left[ \left(
\frac{D-1-2l}{2\omega } \right) ^2+1 \right] , \label{emtRindm01}
\end{eqnarray}
where $l_{m}=D/2-1$ for even $D>2$ and $l_{m}=(D-1)/2$ for odd
$D>1$, and the value for the product over $l$ is equal to 1 for
$D=1,2,3$. These relations illustrate the thermal properties of
the Minkowski vacuum relative to the Rindler observer
\cite{Fulling},\cite{Davies},\cite{Unruh}. As we see from Eq.
(\ref{emtRindm01}), though thermal with temperature $T=(2\pi \xi
)^{-1}$, the spectrum of the vacuum EMT, in general, has a
non-Planckian form: the density of states factor is not
proportional to $\omega ^{D-1} d\omega $. It is interesting to
note that for even values of the space dimension $D $ this thermal
distribution for the scalar field is Fermi-Dirac type.

Note that for the case of a two dimensional spacetime we can not
directly put $D=1$ into the integral (\ref{emtRindm01}). Now,
though the factor $D-1$ in the second term on the right of
expression (\ref{f00}) for $f_0^{(0)}(\omega )$ is equal zero, the
corresponding $\omega $-integral diverges and this term gives
finite contribution to the energy density. Keeping $D>1$ the
$\omega $-integral in Eq. (\ref{emtRindm01}) can be expressed in
terms of the Riemann zeta function $\zeta _{R}(x)$, and in the
limit $D\to 1$ one obtains
\begin{equation}\label{emtRindm01D1}
  \langle T_{0}^{0}\rangle ^{(R)}_{{\rm sub}}=-
  \langle T_{1}^{1}\rangle ^{(R)}_{{\rm sub}}=-
  \frac{1}{\pi \xi ^2}\left(  \frac{1}{24}+\lim _{D\to 1}
  \Gamma (D)\zeta _{R}(D-1)\right) =\frac{1}{2\pi \xi
  ^2}\left( \zeta -\frac{1}{12}\right) ,
\end{equation}
where $\zeta _{R}(0)=-1/2$.

The density of states factor is Planckian for the conformally
coupled scalar and for $D=1,2,3$:
\begin{equation}
\langle T_{i}^{k}\rangle ^{(R)}_{{\rm sub}}=
-\frac{\xi ^{-D-1}}{2^{D-1}\pi ^{D/2}\Gamma (D/2)}%
\int_{0}^{\infty }\frac{\omega ^Dd\omega }{e^{2\pi \omega }+(-1)^D}
{\rm diag}\left( 1,-\frac{1}{D},...,-\frac{1}{D}\right) ,
\quad \zeta =\zeta _c, \quad D=1,2,3.
\label{therm}
\end{equation}
This result for $D=3$ case was obtained by Candelas and Deutsch
\cite{CandD}. Expression (\ref{therm}) corresponds to the absence
from the vacuum of thermal distribution. This means that the
Minkowski vacuum corresponds to a thermal state with respect to
uniformly accelerated observers \cite{Unruh}.

We now treat the case of a massive scalar field. First of all let
us present the difference (\ref{WRMdif}) between Rindler and
Minkowski Wightman functions in another alternative form. For this
we will follow the procedure already used in Ref. \cite{CandRaine}
for the Green function. Substituting \cite{Erdely}
\begin{equation}
K_{i\omega }(\lambda \xi )K_{i\omega }(\lambda \xi ^{\prime })=\frac{1}{2}%
\int_{-\infty }^{+\infty }dy\,e^{i\omega y}K_{0}(\lambda \gamma ),
\label{intK2}
\end{equation}
where $\gamma ^{2}=\xi ^{2}+\xi ^{\prime 2}+2\xi \xi ^{\prime
}\cosh y$, into Eq. (\ref{WRMdif}) and integrating over $\omega $
we receive
\begin{equation}
G_{R}^{+}(x,x^{\prime })-G_{M}^{+}(x,x^{\prime })=-\int \frac{d^{D-1}\mathbf{%
k}}{(2\pi )^{D}}e^{i\mathbf{k}(\mathbf{x}-\mathbf{x}^{\prime
})}\int_{-\infty }^{+\infty }dy\,\frac{K_{0}(\lambda \gamma )}{\pi
^{2}+(y+\tau -\tau ^{\prime })^{2}}.  \label{WRMdif4}
\end{equation}
Integrating over directions of the vector $\mathbf{k}$ with the help of
formula
\begin{equation}
\int d^{D-1}{\mathbf k}\,e^{i\mathbf{k}(\mathbf{x}-\mathbf{x}^{\prime
})}F(k)=(2\pi )^{(D-1)/2}\int_{0}^{\infty }dk\,k^{D-2}F(k)\frac{%
J_{(D-3)/2}(k\left| \mathbf{x}-\mathbf{x}^{\prime }\right| )}{(k\left|
\mathbf{x}-\mathbf{x}^{\prime }\right| )^{(D-3)/2}},  \label{intdirec}
\end{equation}
with $J_{\nu }(x)$ being the Bessel function, and using the value
for the integral over $k$ \cite{Prudnikov2},
\begin{equation}
\int_{0}^{\infty }dk\,k^{\mu +1}J_{\mu }(bk)K_{0}(\gamma \sqrt{k^{2}+m^{2}}%
)=b^{\mu }m^{\mu +1}K_{\mu +1}(m\sqrt{b^{2}+m^{2}}) ,
\label{intkmod}
\end{equation}
one obtains
\begin{equation}
G_{R}^{+}(x,x^{\prime })-G_{M}^{+}(x,x^{\prime })=-\frac{1}{2\pi }%
\int_{-\infty }^{+\infty }\frac{dy}{\pi ^{2}+y^{2}}\left( \frac{m}{2\pi
\gamma _{1}}\right) ^{(D-1)/2}K_{(D-1)/2}(m\gamma _{1}),  \label{WRMdif6}
\end{equation}
where $\gamma _{1}^{2}=\xi ^{2}+\xi ^{\prime 2}+2\xi \xi ^{\prime }\cosh
(y-\tau +\tau ^{\prime })+\left| \mathbf{x}-\mathbf{x}^{\prime }\right| ^{2}$%
. The similar relation for the Feynman's Green function is
considered in Ref. \cite{CandRaine}. Noting that $\gamma _1$ is
just the geodesic separation of the points $x=(\tau ,\xi ,{\mathbf
x})$ and $x^{\prime \prime}=(\tau ^{\prime }+y,-\xi ^{\prime
},{\mathbf x})$ and taking into account the standard expression
for the Minkowskian Wightman function, relation (\ref{WRMdif6})
can be written as
\begin{equation}\label{WRMdif8}
G_{R}^{+}(x,x^{\prime })=G_{M}^{+}(x,x^{\prime })-\int_{-\infty
}^{+\infty }\frac{dy}{\pi ^2+y^2}G_{M}^{+}(x,x^{\prime \prime }).
\end{equation}
Following Ref. \cite{CandRaine}, the second term on the right of
this formula we can interpret as coming from an image charge
density $-1/(\pi ^2+y^2)$ distributed on a line $(-\xi ^{\prime },
{\mathbf x})$ in the left Rindler wedge.

 In the coincidence limit Eq. (\ref{WRMdif6}) yields
\begin{equation}
\langle \varphi ^{2}(x)\rangle _{\mathrm{sub}}^{(R)}=
-\frac{2m^{D-1}}{(2\pi )^{(D+1)/2}}%
\int_{0}^{\infty }\frac{dy}{\pi
^{2}+y^{2}}\frac{K_{(D-1)/2}(z)}{z^{(D-1)/2}},\quad z=2m\xi \cosh
(y/2). \label{phi2dif2}
\end{equation}
This quantity is monotone increasing negative function on $\xi $.
Substituting Eqs. (\ref{WRMdif6}) and (\ref{phi2dif2}) into formula (\ref{vevemtW}%
) for the difference between the corresponding VEV's of the EMT we
obtain (no summation over $i$)
\begin{eqnarray}
\langle T_{0}^{0}\rangle ^{(R)}_{{\rm sub}}&=&
-\frac{2m^{D+1}}{(2\pi )^{(D+1)/2}}\int_{0}^{\infty }\frac{dy}{%
y^{2}+\pi ^{2}}\left[ \frac{K_{(D-1)/2}(z)}{z^{(D-1)/2}}+D\frac{%
K_{(D+1)/2}(z)}{z^{(D+1)/2}}\right] \left( 1-4\zeta \cosh ^{2}(y/2)\right) ,
\label{epsWD1} \\
\langle T_{1}^{1}\rangle ^{(R)}_{{\rm sub}} &=&
\frac{2m^{D+1}}{(2\pi )^{(D+1)/2}}\int_{0}^{\infty }\frac{dy}{%
y^{2}+\pi ^{2}}\frac{K_{(D+1)/2}(z)}{z^{(D+1)/2}}\left( 1-4\zeta \cosh
^{2}(y/2)\right) ,  \label{p1WD} \\
\langle T_{i}^{i}\rangle ^{(R)}_{{\rm sub}} &=&
\frac{2m^{D+1}}{(2\pi )^{(D+1)/2}}\int_{0}^{\infty }\frac{dy}{%
y^{2}+\pi ^{2}}\left\{ \frac{K_{(D+1)/2}(z)}{z^{(D+1)/2}}+\right.
\label{piWD} \\
&& + \left. \left[ \frac{K_{(D-1)/2}(z)}{z^{(D-1)/2}}+(D-1)\frac{%
K_{(D+1)/2}(z)}{z^{(D+1)/2}}\right] (4\zeta -1)\cosh ^{2}(y/2)
\right\} ,\quad i=2,3,\ldots ,D.
\nonumber
\end{eqnarray}
Note that near the horizon, in the limit $m\xi \ll 1$, the leading
terms of the asymptotic expansion for the vacuum EMT components
(\ref{epsWD1})-(\ref{piWD}) do not depend on the mass and coincide
with Eq. (\ref{emtRindm01}). Hence, near the horizon these
components diverge as $\xi ^{-D-1}$. The VEV's of the EMT for the
real massive scalar fields with minimal and conformal couplings on
background of the Rindler spacetime were considered in
Ref.\cite{Hill} by using the covariant functional Schr\"{o}dinger
formalism, developed in Ref. \cite{Hill2}. In this paper for the
EMT the form (\ref{EMT1}) is used with $\zeta =0$ and $\zeta
=\zeta _{c}$ and the corresponding formulas [formulas (2.39) and
(2.45) in Ref.\cite{Hill} ] are more complicated. The equivalency
of these formulas to our expressions (\ref{epsWD1})-(\ref{piWD})
with the special values $\zeta =0,\zeta _{c}$ can be seen using
the identity
\begin{eqnarray}
  \frac{4}{m^2\xi ^{2}}\int_{0}^{\infty }dy \frac{\pi ^2-3y^2}{(\pi ^2+y^2)^3}
  \frac{K_{(D-1)/2}(z)}{z^{(D-1)/2}} &=& \int_{0}^{\infty }\frac{dy}{\pi
  ^2+y^2}\left\{ (1-\cosh y)\frac{K_{(D-1)/2}(z)}{z^{(D-1)/2}}
    \right. \nonumber \\
  & & +\left. \left[ D+1-(D-1)\cosh y\right]\frac{K_{(D+1)/2}(z)}{z^{(D+1)/2}}
  \right\} , \label{IDENTITY2}
\end{eqnarray}
where $z$ is defined as in Eq. (\ref{phi2dif2}). The proof of this
identity we give in Appendix \ref{sec:app3}.

\begin{figure}[tbph]
\begin{center}
\epsfig{figure=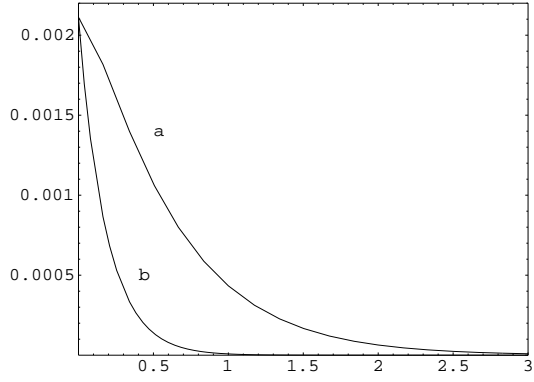,width=7cm,height=5cm}
\end{center}
\caption{ The $D=3$ expectation values for the field square in the
case of the Rindler vacuum without boundaries, $-\xi ^{2}\langle
\varphi ^{2}\rangle _{\mathrm{sub}}^{(R)}$ (curve a) and for the
Minkowski thermal bath, $\xi ^2\langle \varphi ^{2}\rangle
^{\mathrm{th}}_{M}$ (curve b), given by formula
(\ref{Minkthphi2}), versus $m\xi $. } \label{figphi2}
\end{figure}

From expressions (\ref{phi2dif2})-(\ref{piWD}) it follows that for
the massive scalar field the energy spectrum is not strictly
thermal and the quantities $\langle 0_{M}\left| \varphi
^{2}\right| 0_{M}\rangle -\langle 0_{R}\left| \varphi ^{2}\right|
0_{R}\rangle =-\langle \varphi ^{2}(x)\rangle
_{\mathrm{sub}}^{(R)}$ and $\langle 0_{M}\left| T^{k}_{i}\right|
0_{M}\rangle -\langle 0_{R}\left| T^{k}_{i}\right| 0_{R}\rangle
=-\langle T^{k}_{i}(x)\rangle _{\mathrm{sub}}^{(R)}$ do not
coincide with the corresponding ones for the Minkowski thermal
bath, $\langle \varphi ^{2}\rangle ^{\mathrm{th}}_{M}$ and
$\langle T^{k}_{i}\rangle ^{\mathrm{th}}_{M}$, with temperature
$(2\pi \xi )^{-1}$ (for this statement in a variety of situations
see Refs.\cite{Tagaki}-\cite{Medina} and references therein). To
illustrate this for the case $D=3$ we present in Fig.
\ref{figphi2} the quantities $-\langle \varphi ^{2}\rangle
_{\mathrm{sub}}^{(R)}$ (curve a) and
\begin{equation}\label{Minkthphi2}
\langle \varphi ^{2}\rangle ^{\mathrm{th}}_{M}=\frac{1}{2\pi
^2}\int^{\infty }_{0}\frac{\left( \omega ^{2}+2m\omega \right)
^{1/2}d\omega }{e^{(m+\omega )/T}-1}, \quad T=\frac{1}{2\pi \xi }
\end{equation}
(curve b) in dependence of $m\xi $.

\begin{figure}[tbph]
\begin{center}
\begin{tabular}{ccc}
\epsfig{figure=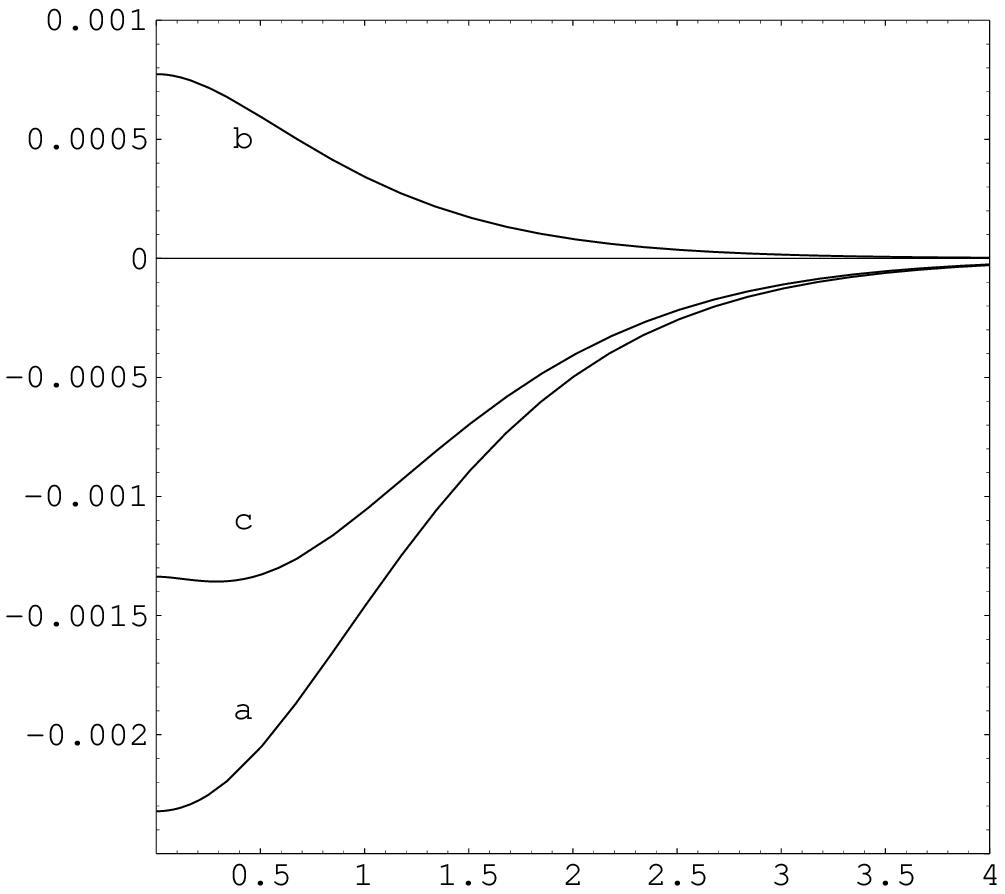,width=7cm,height=6cm} & \hspace*{0.5cm} & %
\epsfig{figure=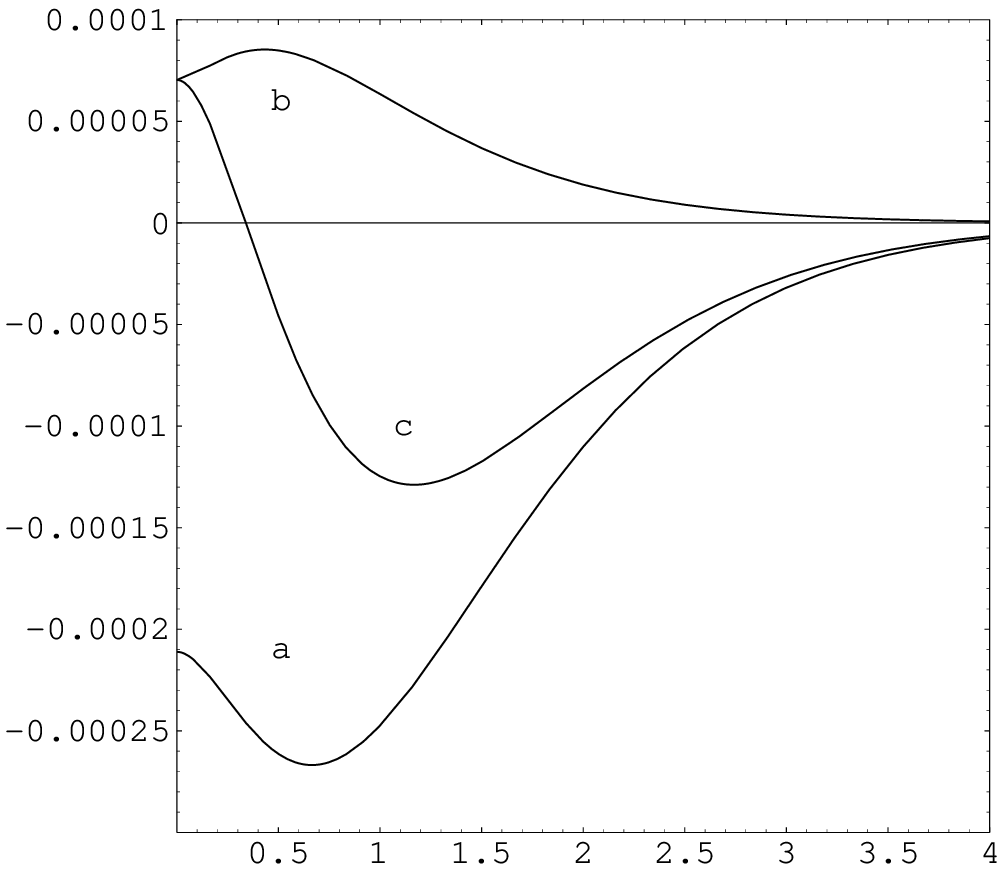,width=7cm,height=6cm}
\end{tabular}
\end{center}
\caption{ The expectation values for the diagonal components of
the energy-momentum tensor, $\xi ^{D+1}\langle T^i_i\rangle
^{(R)}$ (no summation over $i$), versus $m\xi $ in the case of the
Fulling-Rindler vacuum without boundaries for minimally (left) and
conformally (right) coupled scalar fields in $D=3$. The curves a,
b, c correspond to the values $i=0,1,2$ respectively. }
\label{figRind}
\end{figure}

The integrals in formulas (\ref{epsWD1})-(\ref{piWD}) are strongly
convergent and useful for numerical evaluation. In Fig.
\ref{figRind} the results of the corresponding numerical
evaluation for the vacuum EMT components are presented in the
Rindler case without boundaries for minimally and conformally
coupled scalars.

In the limit $m\xi \gg 1$ using the asymptotic formula for the
Macdonald function and taking into account that the main
contribution into the integrals in Eqs. (\ref{epsWD1}),
(\ref{p1WD}), and (\ref{piWD}) comes from the region with small
$y$, in the leading order one finds
\begin{equation}
\langle T_0^0\rangle _{{\rm sub}}^{(R)}\sim \langle T_i^i\rangle
_{{\rm sub}}^{(R)}\sim -2m\xi \langle T_1^1\rangle _{{\rm
sub}}^{(R)}\sim \frac{(4\zeta -1)m^{(D+1)/2}}{2^D\pi ^{(D+3)/2}\xi
^{(D+1)/2}}e^{-2m\xi }, \quad i=2,...,D , \quad m\xi \to \infty .
\label{asympxi}
\end{equation}
In the massless limit, $m=0$, from Eq. (\ref{phi2dif2}) for the
field product one obtains
\begin{equation}
\langle \varphi ^{2}(x)\rangle _{\mathrm{sub}}^{(R)}=
-\frac{\Gamma \left( \frac{D-1}{2}%
\right) }{2^{D}\pi ^{(D+1)/2}\xi ^{D-1}}\int_{0}^{\infty
}\frac{dy}{\pi ^{2}+y^{2}}\cosh ^{1-D}(y/2).  \label{phi2dif2m0}
\end{equation}
The corresponding formulas for the VEV's of the EMT are written as
\begin{eqnarray}
\langle T_{0}^{0}\rangle ^{(R)}_{{\rm sub}} &=&-D \langle
T_{1}^{1}\rangle ^{(R)}_{{\rm sub}}=
\frac{D\Gamma \left( \frac{D+1}{2}\right) }{\left( 2%
\sqrt{\pi }\xi \right) ^{D+1}}\int_{0}^{\infty }\frac{dy}{y^{2}+\pi ^{2}}%
\frac{4\zeta \cosh ^{2}(y/2)-1}{\cosh ^{D+1}(y/2)},  \label{epsWDm0} \\
\langle T_{i}^{i}\rangle ^{(R)}_{{\rm sub}} &=&
\frac{\Gamma \left( \frac{D+1}{2}\right) }{\left( 2\sqrt{\pi }%
\xi \right) ^{D+1}}\int_{0}^{\infty }\frac{dy}{y^{2}+\pi ^{2}}\frac{%
1-(1-4\zeta )(D-1)\cosh ^{2}(y/2)}{\cosh ^{D+1}(y/2)}, \quad
i=2,\ldots ,D. \label{piWDm0}
\end{eqnarray}
Note that the relation (\ref{epsWDm0}) between the energy density
and the vacuum effective pressure also follows from the continuity
equation for the EMT (see Eq. (\ref{conteq}) below). Formulas
(\ref{phi2dif2m0})-(\ref{piWDm0}) are alternative representations
for the VEV's (\ref{phi2Rindm0}), (\ref{emtRindm0}).

\section{VEV's in the RR region}
\label{sec:right}

In Sec. \ref{sec:vev} we have seen that the Wightman function in
the case of the presence of a barrier can be presented in the form
(\ref{WRb}), where the first summand on the right corresponds to
the Fulling-Rindler vacuum without boundaries and the second one
is induced by the barrier and is given by expression (\ref{Wb}).
In the coincidence limit from this formula for the boundary part
of the field square we have
\begin{equation}
\langle \varphi ^2(x)\rangle ^{(b)}=\frac{-1}{2^{D-2}\pi
^{(D+1)/2} \Gamma \left( \frac{D-1}{2}\right) }\int _{0}^{\infty
}dk\, k^{D-2} \int _{0}^{\infty }d\omega \, \frac{\bar I_{\omega
}(\lambda a)}{\bar K_{\omega } (\lambda a)}K_{\omega }^2(\lambda
\xi ),\quad \xi >a. \label{phi2br}
\end{equation}
This quantity is monotone increasing negative function on $\xi $
for Dirichlet scalar and monotone decreasing positive function for
Neumann scalar. Substituting the Wightmann function (\ref{WRb})
into Eq. (\ref{vevemtW}) and taking into account Eqs. (\ref{Wb})
and (\ref{phi2br}) for the VEV's of the EMT in the region $\xi >a$
one finds
\begin{equation}
\langle 0|T_i^k(x)|0\rangle =\langle 0_R|T_i^k(x)|0_R\rangle +
\langle T_i^k(x)\rangle ^{(b)}, \label{emtrig}
\end{equation}
where the first term on the right is the VEV for the
Fulling-Rindler vacuum without boundaries and is investigated in
previous section. The second term is the contribution brought by
the presence of the barrier at $\xi =a$:
\begin{equation}
\langle T_{i}^{k}\rangle ^{(b)}=
\frac{-\delta _{i}^{k}}{2^{D-2}\pi ^{(D+1)/2}\Gamma \left( \frac{D-1}{2}%
\right) }\int_{0}^{\infty }dkk^{D-2}\lambda ^{2}\int_{0}^{\infty
}d\omega \frac{\bar I_{\omega }(\lambda a)}{\bar K_{\omega
}(\lambda a)} F^{(i)}\left[ K_{\omega }(\lambda \xi )\right] ,
\label{emtbound}
\end{equation}
where the functions $F^{(i)}[g(z)]$ , $i=0,1,2$ are obtained from
the functions $f^{(i)}[g(z)]$ in Eqs. (\ref{fq}) replacing $\omega
\rightarrow i\omega $:
\begin{equation}
F^{(i)}[g(z)]=f^{(i)}[g(z),\omega \rightarrow i\omega ].  \label{Ffunc}
\end{equation}
It can be easily checked that both summands on the right of Eq.
(\ref{emtrig}) satisfy the continuity equation $T_{i;k}^{k}=0$,
which for the geometry under consideration takes the form
\begin{equation}
\frac{d}{d\xi }(\xi T_{1}^{1})-T_{0}^{0}=0.  \label{conteq}
\end{equation}

By using the inequalities
\begin{equation}\label{ineqBes}
  I'_{\omega }(z)<\sqrt{1+\omega ^{2}/z^2}I_{\omega }(z),
  \quad -K'_{\omega }(z)>\sqrt{1+\omega ^{2}/z^2}K_{\omega }(z)
\end{equation}
it can be easily seen that $F^{(0)}[K_{\omega }(z)]>0$,
$F^{(1)}[K_{\omega }(z)]<0$ for $\zeta \leq 0$ and
$F^{(i)}[K_{\omega }(z)]>0$, $i=2,3,\ldots ,D$ for $\zeta \leq
\frac{D-3}{4(D-1)}$. In the cases of Dirichlet and Neumann scalars
for the boundary parts of the EMT components these leads
\begin{eqnarray}
\left( \langle T^0_0\rangle ^{(b)}\right) _{{\rm Dirichlet}}&<&0,
\quad  \left( \langle T^1_1\rangle ^{(b)} \right) _{{\rm
Dirichlet}}>0 \label{T00DRR} \\
\left( \langle T^0_0\rangle ^{(b)}\right) _{{\rm Neumann}}&<&0,
\quad \left( \langle T^1_1\rangle ^{(b)} \right) _{{\rm
Neumann}}>0 \label{T00NRR}
\end{eqnarray}
for $\zeta \leq 0$, and (no summation over $i$)
\begin{equation}\label{TiiDNRR}
  \left( \langle T^i_i\rangle ^{(b)}\right) _{{\rm Dirichlet}}<0,
\quad  \left( \langle T^i_i\rangle ^{(b)} \right) _{{\rm
Neumann}}>0, \quad i=2,3,\ldots ,D
\end{equation}
for $\zeta \leq \frac{D-3}{4(D-1)}$.

 For $D=1$ case the corresponding
boundary part is obtained from Eq. (\ref{emtbound})
\begin{equation}
\langle T_{i}^{k}\rangle ^{(b)}=
-\frac{\delta _{i}^{k}}{\pi }m^{2}\,\int_{0}^{\infty }d\omega
\frac{\bar{I}_{\omega }(ma)}{\bar{K}_{\omega }(ma)}F^{(i)}
\left[ K_{\omega }(m\xi )%
\right] ,\quad i=0,1.  \label{D1bound}
\end{equation}
Taking the limit $m\rightarrow 0$ for the massless field we
receive from here
\begin{equation}
\langle T_{i}^{k}\rangle ^{(b)}= \frac{\zeta \delta _{i}^{k}}{\pi
\xi ^{2}}\int_{0}^{\infty }d\omega \,\frac{A+B\omega /a}{A-B\omega
/a}e^{-2\omega \ln (\xi /a)}F_{0}^{(i)},  \label{D1boundm0}
\end{equation}
with
\begin{equation}
F_{0}^{(0)}=2\omega +1,\quad F_{0}^{(1)}=-1.  \label{D1Fi0}
\end{equation}
In this case our assumption for the absence of imaginary
eigenmodes corresponds to the restriction $AB\leq 0$ on the Robin
coefficients. We have obtained result (\ref{D1boundm0}) by using
the analytical continuation from the values $D>1$. If we want to
realize a direct evaluation in $D=1$ the prescription described in
Sec. \ref{sec:vev} has to be modified for the massless case. In
appendix B we show that the corresponding evaluation leads to the
same result.

Boundary part (\ref{emtbound}) is finite for all values $\xi >a$
and diverges at the plate surface $\xi =a$. These surface
divergences are well known in quantum field theory with boundaries
and are investigated for various type of boundary conditions and
geometries. To extract the leading part of the boundary divergence
note that near the boundary the main contribution into the $\omega
$-integral comes from large values of $\omega $ and we can use the
uniform asymptotic expansions for the modified Bessel functions
\cite{abramowiz}. Introducing a new integration variable $k\to
\omega k$ and replacing the Bessel modified functions by their
uniform asymptotic expansions in the leading order one obtains
\begin{eqnarray}
\langle T_0^0(x)\rangle ^{(b)}&\sim &\langle T_2^2(x)\rangle ^{(b)}\sim
\frac{D(2\delta _{B0}-1)(\zeta -\zeta _c)}{2^D\pi ^{(D+1)/2}(\xi -a)^{D+1}}
\Gamma \left( \frac{D+1}{2}\right) , \quad \xi \to a \label{asimpneps} \\
\langle T_1^1(x)\rangle ^{(b)}&\sim & \frac{(1-2\delta
_{B0})(\zeta -\zeta _c)}{2^D\pi ^{(D+1)/2}a(\xi -a)^{D}} \Gamma
\left( \frac{D+1}{2}\right) .  \label{asimpnp}
\end{eqnarray}
These terms do not depend on mass and Robin coefficients, and have
opposite signs for the Dirichlet and Neumann boundary conditions.
Note that the leading terms for $\langle T_0^0(x)\rangle ^{(b)}$
and $\langle T_2^2(x)\rangle ^{(b)}$ are the same as for the plate
in Minkowski spacetime (see, for instance, \cite{RomSah}). This
statement is also valid for the spherical and cylindrical
boundaries on background of the Minkowski spacetime
\cite{Sahsph},\cite{Sahcyl}.

Now let us consider the asymptotic behavior of the boundary part
(\ref{emtbound}) for large $\xi $, $\xi \gg a$. Introducing in Eq.
(\ref{emtbound}) a new integration variable $y=\lambda \xi $ and
using the asymptotic formulas for the Bessel modified functions
for small values of the argument, we see that the subintegrand is
proportional to $(ya/2\xi )^{2\omega }$. It follows from here that
the main contribution into the $\omega $-integral comes from the
small values of $\omega $. Expanding with respect to $\omega $ in
the leading order we obtain
\begin{equation}
\langle T_i^k(x)\rangle ^{(b)}\sim -\frac{\delta _i^k(-1)^{\delta
_{A0}} \xi ^{-D-1}A_{0}^{(i)}(m\xi )}{2^D 3\pi ^{(D-1)/2}(1+\delta
_{A0})\Gamma \left( \frac{D-1}{2}\right) \ln ^2(2\xi /a)} ,
\label{asimpfar}
\end{equation}
where
\begin{equation}
A_{0}^{(i)}(m\xi )=\int _{m\xi }^{\infty }dy\, y^3(y^2-m^2\xi
^2)^{(D-3)/2} F^{(i)}[K_{\omega }(y)]|_{\omega =0} . \label{Ai0}
\end{equation}
If, in addition, one has $m\xi \gg 1$ the integral in this formula
can be evaluated replacing the Macdonald function by its
asymptotic for large values of the argument. In the leading order
this yields
\begin{equation}\label{Aiklargem}
A_{0}^{(0)}\sim A_{0}^{(i)}\sim -2m\xi A_{0}^{(1)}\sim \pi
(1/4-\zeta )\Gamma \left( \frac{D-1}{2}\right) (m\xi
)^{(D+1)/2}e^{-2m\xi }, \quad m\xi \gg 1, \quad i=2,3,\cdots ,D.
\end{equation}
 For the massless case the integral in Eq. (\ref{Ai0}) may
be evaluated using formulas (\ref{IntID})--({\ref{intID2}}) and
one obtains
\begin{equation}
A_0^{(0)}\sim -DA_1^{(1)}\sim \frac{D}{D-1}A_2^{(2)} \sim
\frac{2^DD(\zeta _c-\zeta )}{(D-1)^2\Gamma (D)} \Gamma ^4\left(
\frac{D+1}{2}\right) ,\quad  m=0 . \label{A00}
\end{equation}
For the conformally coupled scalar the leading terms vanish and
the VEV's are proportional to $\xi ^{-D-1}\ln ^{-3}(2\xi /a)$ for
$\xi /a\gg 1$. As it follows from Eqs. (\ref{emtRindm01}),
(\ref{asympxi}), and (\ref{asimpfar}) in both cases
(\ref{Aiklargem}) and (\ref{A00}) and in the region far from the
boundary the VEV's (\ref{emtrig}) are dominated by the purely
Fulling-Rindler part (the first summand on the right): $\langle
T_{i}^{k}\rangle ^{(b)}/\langle T_{i}^{k}\rangle ^{(R)}_{{\rm
sub}}\sim \ln ^{-2}(2\xi /a)$. From Eq. (\ref{asimpfar}) we see
that for a given $\xi $ the boundary part tends to zero as $a \to
0$ (the barrier coincides with the right Rindler horizon) and the
corresponding VEV's of the EMT are the same as for the
Fulling-Rindler vacuum without boundaries. Hence, the barrier
located at the Rindler horizon does not alter the vacuum EMT.

And finally, we turn to the asymptotic $a,\xi \to \infty ,\, \,
\xi -a= {\rm const}$. In this limit $\xi /a\to 1$ and $\omega
$-integrals in Eq. (\ref{emtbound}) are dominated by the large
$\omega $. Replacing the modified Bessel functions by their
uniform asymptotic expansions, keeping the leading terms only and
introducing a new integration variable $\nu =\omega /a $ for the
energy density one obtains
\begin{eqnarray}
\langle T_{0}^{0}\rangle ^{(b)} &\sim & -\frac{1}{2^{D-1}\pi ^{(D+1)/2}
\Gamma \left( \frac{D-1}{2}\right) }\int ^{\infty }_{0}dk \, k^{D-2}
\int_{0}^{\infty }d\nu \frac{e^{-2(\xi -a)\sqrt{\nu ^2+
\lambda ^2}}}{\sqrt{\nu ^2+\lambda ^2}}  \nonumber \\
&& \times \frac{A+B\sqrt{\nu ^2+\lambda ^2}}{A-B\sqrt{\nu
^2+\lambda ^2}} \left[ -4\zeta \nu ^2+\lambda ^2(1-4\zeta )\right]
. \label{asympfl}
\end{eqnarray}
Converting to polar coordinates, $\nu =r\cos \theta ,\, k=r\sin \theta $,
we can integrate over angles with the help of the formula
\begin{equation}
\int _{0}^{\pi /2}d\theta \, \sin ^n\theta =\frac{\sqrt{\pi }}{2}
\frac{\Gamma \left( \frac{n+1}{2}\right)}{\Gamma (n/2+1)} .
\label{intsin}
\end{equation}
Introducing the new integration variable $y=\sqrt{r^2+m^2}$ we receive
\begin{eqnarray}
\langle 0\vert T_0^0\vert 0\rangle &\sim & \langle T_0^0\rangle ^{(b)}
\sim \frac{1}{2^{D}\pi ^{D/2}\Gamma \left( D/2 \right) }\int _{m}
^{\infty }dy\, (y^2-m^2)^{D/2}e^{-2(\xi -a)y} \nonumber \\
&& \times \frac{A+By}{A-By}\left[ 4(\zeta -\zeta _c)+\frac{m^2
(4\zeta -1)}{y^2-m^2}\right],\quad a,\xi \to \infty ,\quad  \xi -
a= {\rm const} . \label{T00ainf}
\end{eqnarray}
Here we have taken into account that in the limit $\xi \to \infty
$ the purely Rindler part $\langle 0_R\vert T_0^0\vert 0_R\rangle
$ vanishes and in Eq. (\ref{emtrig}) the boundary term remains
only. In the same order the evaluation of the components,
corresponding to the vacuum effective pressures in the plane
parallel to the barrier, leads (no summation over $i$)
\begin{equation}
\langle 0\vert T_i^i\vert 0\rangle \sim \langle 0\vert T_0^0\vert
0\rangle , \quad i=2,3,...,D, \label{Tiiainf}
\end{equation}
The leading term for the component $\langle 0\vert T_1^1\vert
0\rangle $ vanishes,
\begin{equation}
\langle 0\vert T_1^1\vert 0\rangle ={\mathcal O}(1/a),\quad a\to \infty .
\label{T11ainf}
\end{equation}
The term to the next order can be found using continuity equation
(\ref{conteq}) and expression (\ref{T00ainf}). The expectation
values (\ref{T00ainf}), (\ref{Tiiainf}), and (\ref{T11ainf})
coincide with the corresponding VEV's induced by a single plate in
the Minkowski spacetime with the Robin boundary conditions on it,
and are investigated in Ref. \cite{RomSah} for the massless field
and in Ref. \cite{MatSah} for the massive scalar field.

\section{Vacuum stresses in the RL region}
\label{sec:left}

\subsection{Scalar field}

In previous section we have considered VEV's for the EMT in the region $%
\xi >a$. Now we will turn to the case of the scalar field in
region between the barrier and Rindler horizon (RL region), $0<\xi
<a$, and satisfying boundary condition (\ref{boundRind}) on the
boundary $\xi =a$. For this region we have to take both solutions
to equation (\ref{eqxi}) and the eigenfunctions have the form
(\ref{sol1}), where now
\begin{equation}
\phi (\xi )=Z_{i\omega }(\lambda \xi ,\lambda a)\equiv K_{i\omega }(\lambda
\xi )-\frac{\bar K_{i\omega }(\lambda a)}{\bar I_{i\omega }
(\lambda a)}I_{i\omega }(\lambda \xi ),  \label{Zet}
\end{equation}
and the coefficient is obtained from boundary condition
(\ref{boundRind}). Here the quantities with overbars are defined
in accordance with Eq. (\ref{barnot}). As we see, unlike to the
previous case, the spectrum for $\omega $ is continous. The
normalization coefficient is determined from condition
(\ref{normcond}), where integration over $\xi $ goes between the
limits $(0,a)$. Using integration formula (\ref{intformgen}) and
the representation for the delta function
\begin{equation}
\delta (x)=\lim_{\sigma \to \infty }\frac{\sin \sigma x}{\pi x} ,
\label{delta}
\end{equation}
it can be seen that
\begin{equation}
C=\frac{\sqrt{\sinh \omega \pi }}{\pi (2\pi )^{(D-1)/2}}.  \label{coef2}
\end{equation}
Substituting eigenfunctions (\ref{sol1}) with the function
(\ref{Zet}) into the mode sum for the Wightman function one finds
\begin{equation}
G^{+}(x,x')=
 \frac{1}{\pi ^2}\int \frac{d^{D-1}{\mathbf k}}{(2\pi )^{D-1}}\,
e^{i{\mathbf k}({\mathbf x}-{\mathbf x'})}\int_{0}^{\infty }d\omega \sinh (\pi
\omega ) e^{-i\omega (\tau -\tau ')}Z_{i\omega }(\lambda \xi ,
\lambda a)Z_{i\omega }^{*}(\lambda \xi ',\lambda a) .  \label{Wleft}
\end{equation}
The part induced by the barrier can be obtained subtracting from
here the Wightman function for the right Rindler wedge without
boundaries, given by expression (\ref{emtRindler}). To transform
the corresponding difference note that
\begin{equation}
Z_{i\omega }(\lambda \xi ,\lambda a)Z_{i\omega }^{*}
(\lambda \xi ',\lambda a)-K_{i\omega }(\lambda \xi )
K_{i\omega }^{*}(\lambda \xi ')=\frac{\pi \bar K_{i\omega }(\lambda a)}{2i
\sinh \pi \omega }%
\sum_{\sigma =-1,1}\sigma \frac{I_{i\sigma \omega }(\lambda \xi )
I_{i\sigma \omega }(\lambda \xi ')}{%
\bar I_{i\sigma \omega }(\lambda a)},  \label{difZK}
\end{equation}
where we have used the standard relation
\begin{equation}
K_{\nu }(z)=\frac{\pi }{2\sin \nu \pi }\left[ I_{-\nu
}(z)-I_{\nu }(z)\right] .  \label{relNKI}
\end{equation}
This allows to present the boundary part in the form
\begin{eqnarray}
\langle \varphi (x)\varphi (x')\rangle ^{(b)} &=&
G^{+}(x,x')-G^{+}_{R}(x,x') \nonumber \\
&= & \frac{1}{i}\int \frac{d^{D-1}{\mathbf k}}{(2\pi )^{D}}\,
e^{i{\mathbf k}({\mathbf x}-{\mathbf x'})}\int_{0}^{\infty }d\omega
 e^{-i\omega (\tau -\tau ')}\sum_{\sigma =-1,1}\sigma
\frac{\bar K_{i \omega }(\lambda a)}{\bar I_{i \sigma \omega
}(\lambda a)} I_{i\sigma \omega }(\lambda \xi )I_{i\sigma \omega
}(\lambda \xi '). \label{Wbleft1}
\end{eqnarray}
Assuming that the function $\bar I_{i\omega }(\lambda a)$ ($\bar
I_{-i\omega }(\lambda a)$) has no zeros for $-\pi /2\leq {\rm
arg}\, \omega <0$ ($0< {\rm arg}\, \omega <\pi /2$) we can rotate
the integration contour over $\omega $ by angle $-\pi /2$ for the
term with $\sigma =1$ and by angle $\pi /2$ for the term with
$\sigma =-1 $. The integrals taken around the arcs of large radius
tend to zero under the condition $|\xi \xi '|<a^2e^{|\tau -\tau
'|}$ (note that, in particular, this is the case in the
coincidence limit for the region under consideration). As a result
for the difference of the Wightman functions one obtains
\begin{equation}
\langle \varphi (x)\varphi (x')\rangle ^{(b)} =
-\frac{1}{\pi }\int \frac{d^{D-1}{\mathbf k}}{(2\pi )^{D-1}}\,
e^{i{\mathbf k}({\mathbf x}-{\mathbf x'})}\int_{0}^{\infty }d\omega
\frac{\bar K_{\omega }(\lambda a)}{\bar I_{\omega }(\lambda a)}
I_{\omega }(\lambda \xi )I_{\omega }(\lambda \xi ')
\cosh \left[ \omega (\tau -\tau ')\right] . \label{Wbleft2}
\end{equation}
As we see this expression differs from the corresponding formula
(\ref{Wb}) for the region $a<\xi <\infty $ by replacements
$I_{\omega }\rightarrow K_{\omega }$, $K_{\omega }\rightarrow
I_{\omega }$. In the coincidence limit for the VEV of the field
square we receive
\begin{equation}
\langle \varphi ^2(x)\rangle ^{(b)} =
\frac{-1}{2^{D-2}\pi ^{(D+1)/2}\Gamma \left( \frac{D-1}{2}\right) }
\int _{0}^{\infty }dk\, k^{D-2}\int_{0}^{\infty }d\omega
\frac{\bar K_{\omega }(\lambda a)}{\bar I_{\omega }(\lambda a)}
I_{\omega }^{2}(\lambda \xi ),\quad 0<\xi <a .
\label{phi2bl}
\end{equation}
This quantity is monotone decreasing negative function on $\xi $
for Dirichlet scalar and monotone increasing positive function for
Neumann scalar. From Eq. (\ref{vevemtW}) the VEV's of the EMT are
obtained in the form (\ref{emtrig}) with the boundary part
\begin{equation}
\langle T_{i}^{k}\rangle ^{(b)}=-\frac{\delta _{i}^{k}}{2^{D-2}
\pi ^{(D+1)/2}\Gamma \left( \frac{D-1%
}{2}\right) }\int_{0}^{\infty }dk\,k^{D-2}\lambda ^{2}\int_{0}^{\infty
}d\omega \,\frac{\bar K_{\omega }(\lambda a)}{\bar I_{\omega }(\lambda a)}
F^{(i)} \left[ I_{\omega }(\lambda \xi )\right] ,  \label{emtboundN1}
\end{equation}
where, as in Eq. (\ref{emtbound}), the functions $F^{(i)}[g(z)]$
are obtained from expressions (\ref{fq}) replacing $\omega \to
i\omega $ (see Eq. (\ref{Ffunc})). Comparing Eqs. (\ref{phi2bl}),
(\ref{emtboundN1}), (\ref{phi2br}), and (\ref{emtbound}) we see
that the boundary parts for the region $0<\xi <a$ can be obtained
from the corresponding ones for the region $\xi >a$ by the
replacements $I_{\omega }\to K_{\omega }$, $K_{\omega }\to
I_{\omega }$. Note that here the situation is the same as for the
interior and exterior regions in the case of cylindrical and
spherical surfaces on background of the Minkowski spacetime (see,
\cite{Sahsph},\cite{Sahcyl}). The reason for this analogy is that
the uniformly accelerated observers produce world lines in
Minkowski spacetime which in Euclidean space would correspond to
circles. This correspondence extends also to the field equation
and its solutions, and modes (\ref{sol1}) are the Minkowski
spacetime analogue corresponding to cylinder harmonics.

For $D=1$ case the integral over ${\mathbf k}$ in Eq.
(\ref{Wbleft2}) is absent and $\lambda =m$. The corresponding
formulas for the field square and vacuum EMT are obtained from
Eqs. (\ref{phi2bl}) and (\ref{emtboundN1}) replacing
\begin{equation}
\frac{1}{2^{D-2}
\pi ^{(D+1)/2}\Gamma \left( \frac{D-1%
}{2}\right) }\int_{0}^{\infty }dk\,k^{D-2} \to \frac{1}{\pi },
\quad \lambda \to m.
\label{D1repl}
\end{equation}
In particular, for the massless $D=1$ case we have
\begin{equation}
\langle T_i^k\rangle ^{(b)}=\frac{\zeta \delta _i^k}{\pi \xi ^2}
\int _{0}^{\infty }d\omega \, \frac{A-B\omega /a}{A+B\omega /a}
e^{-2\omega \ln (a/\xi )}\tilde F_0^{(i)},
\label{D1m0left}
\end{equation}
where
\begin{equation}
\tilde F_0^{(0)}=2\omega -1,\quad \tilde F_0^{(1)}=1.
\label{Fi0tild}
\end{equation}

The expressions (\ref{emtboundN1}) are divergent on the plate
surface, $\xi =a$. The leading terms for the corresponding
asymptotic expansions are derived by the way similar to that for
Eqs. (\ref{asimpneps}) and (\ref{asimpnp}), and can be obtained
from these formulas replacing $\xi -a$ by $a-\xi $ and changing
the sign for $\langle T_1^1\rangle ^{(b)}$.

Now let us consider the behavior of the boundary part
(\ref{emtboundN1}) for $\xi \ll a$. Introducing instead of $k$ a
new integration variable $y=\lambda a$ and using the formula for
$I_{\omega }(y\xi /a)$ in the case of small values of the argument
we see that the $\omega $ - subintegrand is proportional to
$e^{-2\omega \ln (a/y\xi )}$. It follows from here that the main
contribution comes from the small values of $\omega $. Expanding
over $\omega $ we obtain the standard integrals of the form $\int
_{0}^{\infty }\omega ^ne^{-2\omega \ln (a/y\xi )}d\omega $.
Performing these integrals one obtains in the leading order
\begin{eqnarray}
\langle T_0^0\rangle ^{(b)}&\sim &-\langle T_1^1\rangle ^{(b)}\sim -
\frac{\zeta B_0(ma)\left( 1+{\mathcal O}(\ln ^{-1}(2a/\xi ))
\right) }{2^{D-1}\pi ^{(D+1)/2}\Gamma \left(\frac{D-1}{2}\right)
a^{D-1}\xi ^{2}\ln ^{2}(2a/\xi )}  \label{asimpxi0}\\
\langle T_i^i\rangle ^{(b)}&\sim & \frac{(4\zeta -1)B_0(ma)\left(
1+{\mathcal O}(\ln ^{-1}(2a/\xi )) \right) }{2^{D}\pi
^{(D+1)/2}\Gamma \left(\frac{D-1}{2}\right) a^{D-1}\xi ^{2}\ln
^{3}(2a/\xi )} ,\quad i=2,3,...,D, \label{asimpxi0i}
\end{eqnarray}
where
\begin{equation}
B_0(ma)=\int_{ma}^{\infty }dy\, y(y^2-m^2a^2)^{(D-3)/2}\frac{\bar
K_0 (y)}{\bar I_0(y)}. \label{asimpB0}
\end{equation}
Note that the function $B_0(ma)$ is positive for Dirichlet
boundary condition and is negative for Neumann one. For $D=1$ case
one has
\begin{equation}
\langle T_0^0\rangle ^{(b)} \sim -\langle T_1^1\rangle ^{(b)}\sim
- \frac{\zeta }{2\pi \xi ^2\ln ^{2}(2/m\xi )}\frac{\bar K_0
(ma)}{\bar I_0(ma)} . \label{asimpxi0D1}
\end{equation}
As we see the boundary part is divergent at Rindler horizon.
Recall that near the horizon the purely Rindler part behaves as
$\xi ^{-D-1}$ and, therefore, in this limit the total vacuum EMT
is dominated by this part.

\subsection{Electromagnetic field}

We now turn to the case of the electromagnetic field in the region
$0<\xi <a$ for the case $D=3$. We will assume that the mirror is a
perfect conductor with the standard boundary conditions of
vanishing of the normal component of the magnetic field and the
tangential components of the electric field, evaluated at the
local inertial frame in which the conductor is instantaneously at
rest. As it has been shown in Ref. \cite{CandD} the corresponding
eigenfunctions for the vector potential $A_{\mu }(x)$ may be
resolved into transverse electric (TE) and transverse magnetic
(TM) (with respect to $\xi $-direction) modes:
\begin{equation}
A_{\mu }^{(\sigma )}(x)=\left\{
\begin{array}{cc}
(0,0,-ik_{3},ik_{2})\varphi ^{(0)}(x),
\quad  & \mathrm{for}\quad \sigma =0
\\
(-\xi \partial /\partial \xi ,i\omega /\xi ,0,0)\varphi ^{(1)}(x),
\quad  &
\mathrm{for}\quad \sigma =1
\end{array}
\right. ,  \label{eigel}
\end{equation}
where ${\mathbf k}=(k_{2},k_{3})$, $\sigma =0$ and $\sigma =1$ correspond to
the TE and TM waves respectively. From the perfect conductor boundary
conditions on the vector potential we obtain the corresponding boundary
conditions for the scalar fields $\varphi ^{(\sigma )}(x)$:
\begin{equation}
\varphi ^{(0)}(x)=0,\quad \frac{\partial \varphi ^{(1)}(x)}{\partial \xi }%
=0,\quad \xi =a.  \label{elbound}
\end{equation}
As a result the TE/TM modes correspond to the Dirichlet/Neumann scalars.
In the corresponding expressions for the eigenfunctions
$A_{\alpha \mu }^{(\sigma )}(x)$ the normalization
coefficient is determined from the orthonormality condition
\begin{equation}
\int d{\mathbf x}\int _{0}^{a}\frac{d\xi }{\xi }
A_{\alpha }^{(\sigma )\mu }(x)A_{\alpha '\mu }^{(\sigma ')* }(x)
=-\frac{2\pi }{\omega }\delta _{\alpha \alpha '}\delta _{\sigma \sigma '},
\quad \alpha =({\mathbf k},\omega ).
\label{normal}
\end{equation}
From here for the corresponding Diriclet
and Neumann modes we have
\begin{equation}
\varphi ^{(\sigma )}_{\alpha }(x)=
\frac{\sqrt{\sinh \omega \pi }}{\pi ^{3/2}k}
\left[ K_{i\omega }(k\xi )-\frac{K_{i\omega }^{(\sigma )}(ka)}{%
I_{i\omega }^{(\sigma )}(ka)}I_{i\omega }(k\xi )\right] e^{i{\mathbf kx}%
-i\omega \tau },  \label{elsceig}
\end{equation}
where $K_{i\omega }^{(0)}(z)=K_{i\omega }(z)$, $K_{i\omega
}^{(1)}(z)=dK_{i\omega }(z)/dz$ and the same notations for the
function $I_{i\omega }(z)$. The VEV's for the EMT can be obtained
substituting eigenfunctions (\ref{eigel}) into the mode sum
formula
\begin{equation}
\langle 0\left| T_{i}^{k}(x)\right| 0\rangle =
\sum_{\sigma =0,1}\int d{\mathbf k}%
\int_{0}^{\infty }d\omega \,T_{i}^{k}\left\{ A_{\alpha \mu }^{(\sigma )}(x),
A_{\alpha \mu
}^{(\sigma )\ast }(x)\right\} ,  \label{vevel}
\end{equation}
with the standard bilinear form for the electromagnetic field EMT:
\begin{equation}
T_i^k\left\{ A_{\mu }(x),A_{\mu }(x)\right\}=\frac{1}{4\pi }
\left( -F_{il}F^{lk}+\frac{1}{4}\delta _i^k
F_{ln}F^{ln}\right) ,
\label{elemt}
\end{equation}
where $F_{lk}=\partial A_k/\partial x^l-\partial A_l/\partial x^k$
is the field tensor. Substituting eigenfunctions (\ref{eigel})
into the mode sum (\ref{vevel}) one finds
\begin{equation}
\langle 0\left| T_{i}^{k}(x)\right| 0\rangle =
\frac{\delta _{i}^{k}}{4\pi ^{3}}\sum_{\sigma
=0,1}\int _{0}^{\infty }dk\,k^{3}\int_{0}^{\infty }
d\omega \,\sinh (\omega \pi )f_{\mathrm{%
em}}^{(i)}\left[ K_{i\omega }(k\xi )-\frac{K_{i\omega }^{(\sigma )}(ka)}{%
I_{i\omega }^{(\sigma )}(ka)}I_{i\omega }(k\xi )\right] .  \label{emtel}
\end{equation}
Here for a given function $g(z)$ the following notations are introduced
\begin{eqnarray}
f_{\mathrm{em}}^{(0)}\left[ g(z)\right]  &=&\left| \frac{dg(z)}{dz}\right|
^{2}+\left( 1+\frac{\omega ^{2}}{z^{2}}\right) \left| g(z)\right| ^{2},
\nonumber \\
f_{\mathrm{em}}^{(1)}\left[ g(z)\right]  &=&-\left| \frac{dg(z)}{dz}\right|
^{2}+\left( 1-\frac{\omega ^{2}}{z^{2}}\right) \left| g(z)\right| ^{2},
\label{fiel} \\
f_{\mathrm{em}}^{(2)}\left[ g(z)\right]  &=&f_{\mathrm{em}}^{(3)}\left[ g(z)%
\right] =-\left| g(z)\right| ^{2}.  \nonumber
\end{eqnarray}
It can be easily checked that components (\ref{emtel}) satisfy
zero trace condition and continuity equation (\ref{conteq}). The
way for subtraction from these quantities the parts due to the
Fulling-Rindler vacuum without boundaries is the same as for the
scalar field, given above in this section. This yields to the
following result
\begin{equation}
\langle 0\left| T_{i}^{k}\right| 0\rangle =\langle 0_{R}\left| T_{i}^{k}
\right| 0_{R}\rangle -\frac{\delta
_{i}^{k}}{4\pi ^{2}}\int _{0}^{\infty }dk\,k^{3}\int_{0}^{\infty }
d\omega \,\left[ \frac{%
K_{\omega }(ka)}{I_{\omega }(ka)}+\frac{K_{\omega }^{\prime }(ka)}{I_{\omega
}^{\prime }(ka)}\right] F_{\mathrm{em}}^{(i)}\left[ I_{\omega }(k\xi )\right]
,  \label{emtboundel}
\end{equation}
where
\begin{equation}
\langle 0_{R}\left| T_{i}^{k}\right| 0_{R}\rangle =\langle 0_{M}
\left| T_{i}^{k}\right| 0_{M}\rangle -\frac{1%
}{\pi ^{2}\xi ^{4}}\int_{0}^{\infty }\frac{d\omega \,(\omega ^{3}+\omega )}{%
e^{2\pi \omega }-1}\mathrm{diag}\left( 1,-\frac{1}{3},-\frac{1}{3},-\frac{1}{%
3}\right)   \label{Rindel}
\end{equation}
are the VEV's for the Fulling-Rindler vacuum without boundaries
\cite{CandD}, and the functions $F_{\mathrm{em}}^{(i)}\left[
g(z)\right] $ are obtained from Eq. (\ref{fiel}) replacing
$\omega \rightarrow i\omega $: $F_{\mathrm{em}}^{(i)}%
\left[ g(z)\right] =f_{\mathrm{em}}^{(i)}\left[ g(z),\omega
\rightarrow i\omega \right] $. Note that the $\omega $-integral in
Eq. (\ref{Rindel}) is equal to 11/240. Comparing the boundary part
in Eq. (\ref{emtboundel}) to the corresponding expression for the
RR region, $\xi >a$, derived in \cite{CandD}, we see that, as in
the scalar case, these results are obtained from each other by the
replacements $I_\omega \to K_\omega$, $K_\omega \to I_\omega$.

By using inequalities (\ref{ineqBes}) we conclude that
\begin{equation}\label{ineqBes2}
  \frac{K_{\omega }(z)}{I_{\omega }(z)}+
  \frac{K'_{\omega }(z)}{I'_{\omega }(z)}<0.
\end{equation}
This, together with the observation that $I_{\omega }(z)$ is
monotone increasing function, immediately shows that the
components $\langle T_{i}^{i}\rangle ^{(b)}$, $i=2,3$ are negative
and monotone decreasing on $\xi $. Further from the first
inequality (\ref{ineqBes}) one has $F^{(1)}_{{\rm em}}[I_{\omega }
(k\xi ) ]>0$ and it can be seen that this function is monotone
increasing on $\xi $. Hence, the component $\langle
T_{1}^{1}\rangle ^{(b)}$ is positive and monotone increasing. And,
at last, by using Eq. (\ref{ineqBes}) we see that $F^{(0)}_{{\rm
em}}[I_{\omega } (k\xi ) ]>0$, and as a result one obtains that
the boundary part of the energy density is positive, $\langle
T_{0}^{0}\rangle ^{(b)}>0$.

Now let us consider the asymptotics of the boundary part in Eq.
(\ref{emtboundel}) near the barrier, $\xi \to a$. In this limit
the corresponding expressions are divergent. It follows from here
that the main contributions into the VEV's are due to the large
$\omega $. Rescaling the integration variable $k\to \omega k$ and
replacing the Bessel modified functions by their uniform
asymptotic expansions in the leading order one finds
\begin{eqnarray}
\langle T_{0}^{0}\rangle ^{(b)} & \sim & -2\langle T_{i}^{i}\rangle ^{(b)}
\sim \frac{1}{30\pi ^2a(a-\xi )^3},\quad i=2,3,  \label{asimpela} \\
\langle T_{1}^{1}\rangle ^{(b)} & \sim & \frac{1}{60\pi
^2a^2(a-\xi )^2}, \quad \xi \to a-0 .\nonumber
\end{eqnarray}
Note that the corresponding asymptotics for the region $\xi >a$
are given by the same formulas \cite{CandD}.

Now we turn to the asymptotics in the limit $\xi \to 0$. Replacing
the function $I_{\omega }(k\xi )$ by its asymptotic, and noting
that the main contribution into the $\omega $-integrals come from
the region with small $\omega $, in the leading order we have (no
summation over $i$)
\begin{equation}
\langle T_{0}^{0}\rangle ^{(b)} \sim \langle T_{1}^{1}\rangle
^{(b)} \sim -\langle T_{i}^{i}\rangle ^{(b)}\sim -\frac{1}{8\pi
^2a^4\ln (2a/\xi )} \int _{0}^{\infty }dy\, y^3\left[
\frac{K_0(y)}{I_0(y)}- \frac{K_1(y)}{I_1(y)}\right] .
\label{asimpel0}
\end{equation}
In this formula the value for the integral is equal to -1.326. As
we see, unlike to the case for the scalar field, the boundary part
of the electromagnetic vacuum EMT tends to zero as $\xi \to 0$.

And finally, let us consider the leading terms in the vacuum EMT
in the limit $a,\xi \to \infty $, $\xi -a ={\rm constant}$.
Introducing in Eq. (\ref{emtboundel}) a new integration variable
$ka$ we see that in this limit the corresponding terms have the
same form as in the case $\xi \to a$, $a={\rm constant}$ and are
given by expressions (\ref{asimpela}). Note that in this limit the
total vacuum EMT is dominated by the boundary part.

\begin{figure}[tbph]
\begin{center}
\begin{tabular}{ccc}
\epsfig{figure=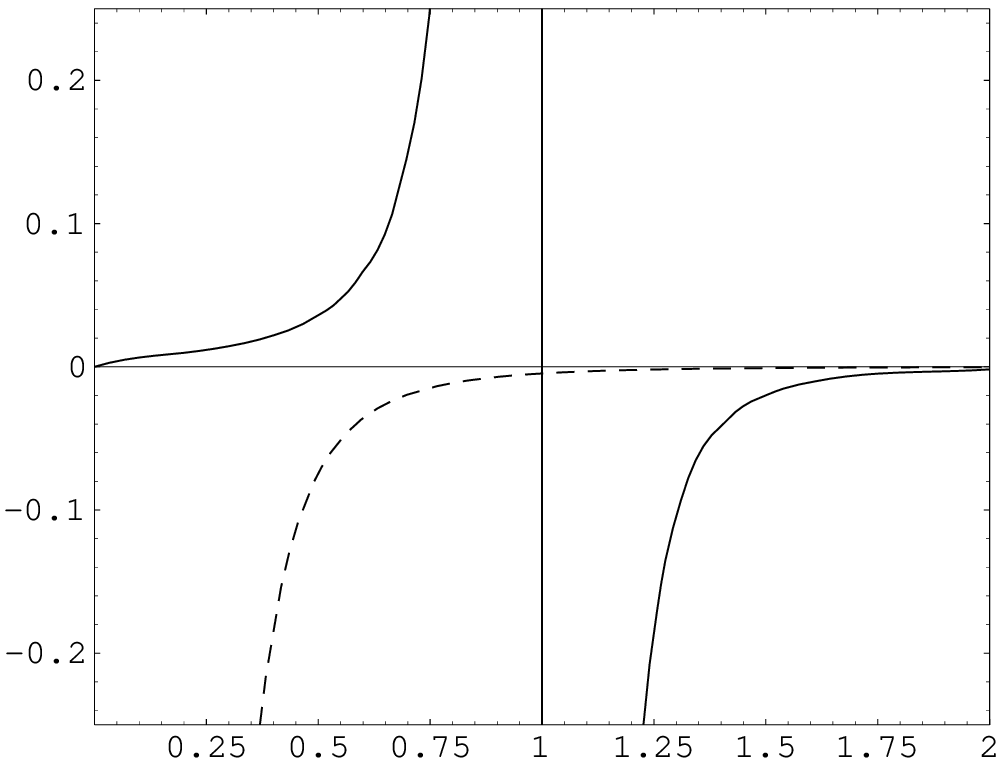,width=7cm,height=6cm} & \hspace*{0.5cm} & %
\epsfig{figure=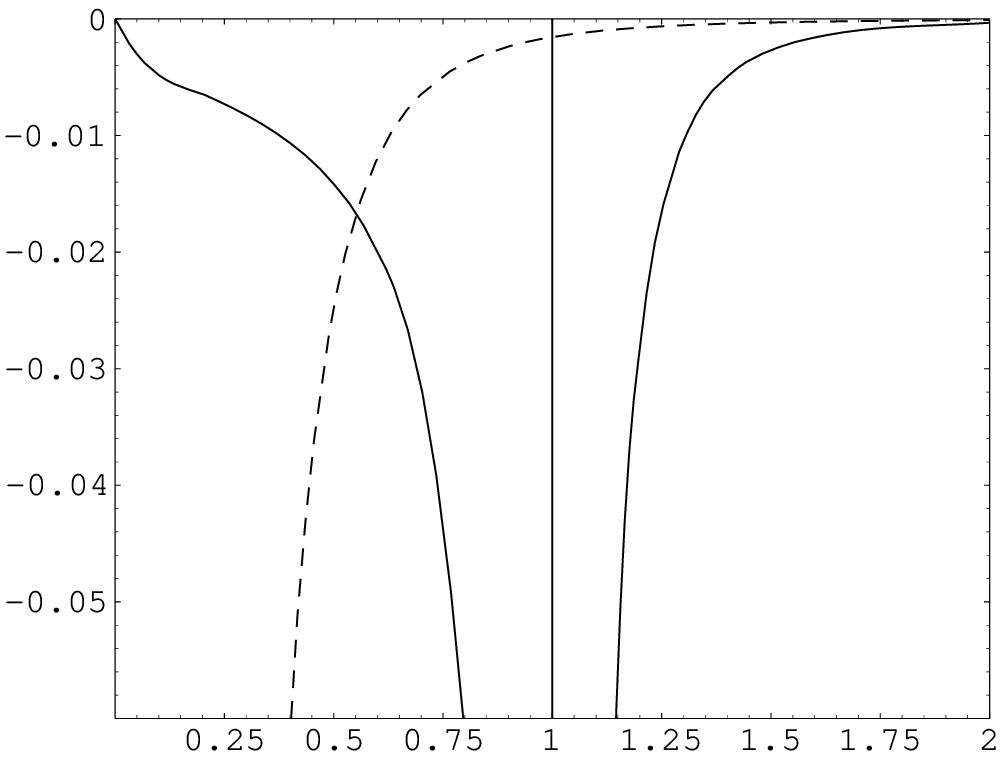,width=7cm,height=6cm}
\end{tabular}
\end{center}
\caption{ The expectation values for the boundary part of the
energy density, $a^4\langle T_{0}^{0}\rangle ^{(b)}$ (left), and
perpendicular effective pressure, $-a^4\langle T_{1}^{1}\rangle
^{(b)}$ (right), multiplied by $a^4$, for the electromagnetic
field versus $\xi /a$. The dashed curves present the corresponding
regularized quantities for the Fulling-Rindler vacuum without
boundaries [see Eq. (\ref{Rindel})]. } \label{fig2el}
\end{figure}

In  Fig. \ref{fig2el} we have plotted the electromagnetic vacuum
energy density, $a^4\langle T^0_0\rangle $ (left), and
perpendicular pressure, $-a^4\langle T^1_1\rangle $ (right), in
dependence of $\xi /a$, generated by a single perfectly conducting
plate uniformly accelerated normal to itself. For the plate $\xi
=a$. The dashed curves present the corresponding regularized
quantities for the Fulling-Rindler vacuum without boundaries,
given by the second summand on the right of formula
(\ref{Rindel}). The full curves are for the boundary parts, second
term on the right of formula (\ref{emtboundel}). In addition to
the region $0<\xi <a$, investigated in this section, we have also
included the region $a<\xi <\infty $. The formulas for the
boundary part in the latter case are derived in Ref. \cite{CandD}
and, as it have been mentioned above, are obtained from the ones
considered here by replacements $I_\omega \to K_\omega$, $K_\omega
\to I_\omega $. From Fig. \ref{fig2el} we see that the vacuum
perpendicular pressure is always negative.
The total energy density (as a sum of purely Rindler and boundary
parts) is always negative in the region $a<\xi <\infty $. For the
region $0<\xi < \infty $ the total energy density is negative near
the horizon and is positive near the plate. As a result for some
intermediate value of $\xi $ the energy density vanishes.

\section{Conclusions} \label{sec:conc}

Quantum field theory in accelerated systems contains many of the
special features produced by a gravitational field without the
complications due to a curved background spacetime. In this paper
we have considered the VEV's of the EMT induced by uniformly
accelerated plane boundary in the Fulling-Rindler vacuum. This
problem for the region on the right of the plane has been
previously studied by Candelas and Deutch in the paper
\cite{CandD}. These authors consider a conformally coupled scalar
field with Dirichlet or Neumann boundary conditions and
electromagnetic field. Here we generalize the corresponding
results for (i) an arbitrary number of spatial dimension, (ii) for
a scalar field with general curvature coupling parameter and (iii)
nonzero mass, (iv) for mixed boundary conditions of Robin type. In
addition, (v) we consider the VEV's in the region between the
plane and Rindler horizon for scalar and electromagnetic fields.

To obtain the expectation values for the energy-momentum tensor we
first construct the positive frequency Wightman function (this
function is also important in considerations of the response of a
particle detector \cite{Birrel}). For the region on the right of
the plate, $\xi
>a$, the application of the generalized Abel-Plana formula to the
mode sum over zeros of the function $\bar K_{i\omega }$ allows us
to extract the purely Rindler part without boundaries and to
present the additional boundary part in terms of strongly
convergent integrals (formulas (\ref{WRb}), (\ref{Wb})). The
expectation values for the EMT are obtained by applying a certain
second order differential operator and taking the coincidence
limit. First, in Sec. \ref{sec:Rindler}, we consider the vacuum
EMT in the Rindler wedge without boundaries. After the subtraction
of the Minkowskian part the corresponding components are presented
in the form (\ref{emtRsub}) with notations (\ref{fq}). For the
massless scalar the VEV's of the EMT take the form
(\ref{emtRindm01}). This expression corresponds to the absence
from the vacuum of thermal distribution with standard temperature
$T=(2\pi\xi)^{-1}$. In general, this distribution has
non-Planckian spectrum: the density of states factor is not
proportional to $\omega ^{D-1}d\omega $. The spectrum takes the
Planckian form for conformally coupled scalars in $D=1,2,3$. It is
interesting to note that for even values of spatial dimension the
distribution is Fermi-Dirac type. Further, on the base of
procedure already used in \cite{CandRaine}, for the massive case
we present the vacuum EMT in another alternative form, formulas
(\ref{epsWD1})-(\ref{piWD}). By using the identity proved in
Appendix \ref{sec:app3} we show that for the cases of conformally
and minimally coupled scalars these formulas are equivalent to
those previously obtained in Ref. \cite{Hill}. For the massive
scalar the energy spectrum is not strictly thermal and the
corresponding quantities do not coincide with ones for the
Minkowski thermal bath with temperature $(2\pi\xi)^{-1}$. This
fact is illustrated in Fig. \ref{figphi2}. In the limit $m\xi \gg
1$ the vacuum EMT components are exponentially decreasing
functions on $m\xi $. In the massless limit we obtain another
equivalent representations, Eqs. (\ref{epsWDm0}) and
(\ref{piWDm0}).

In Sec. \ref{sec:right} we investigate the boundary part of the
vacuum EMT induced by a single plate in the region $\xi >a$,
formula (\ref{emtbound}). The corresponding result for the
massless 1D field can be obtained by the analytic continuation and
has form (\ref{D1boundm0}). In Appendix \ref{sec:app2} we show
that this result coincides with the formula derived by direct
calculation in $D=1$. Further we investigate the asymptotic
properties of the boundary EMT. Near the plate surface the total
VEV's are dominated by the boundary part and the corresponding
components diverge at the boundary. For non-conformally coupled
scalars the leading terms are given by formulas (\ref{asimpneps})
and (\ref{asimpnp}) and are the same as for an infinite plane
boundary in Minkowski spacetime with Dirichlet and Neumann
boundary conditions. These terms do not depend on the mass or
Robin coefficients, and have opposite signs for Dirichlet and
Neumann cases. For large values of $\xi\gg a$ the leading term in
the asymptotic expansion of the boundary induced EMT does not
depend on the Robin coefficient $B$ and has the form
(\ref{asimpfar}). In this limit $\langle T_i^k\rangle
^{(b)}/\langle T_i^k\rangle ^{(R)}_{{\rm sub}}\sim \ln ^{-2}(2\xi
/a) $ and the vacuum EMT is dominated by the purely Rindler part.
For a given $\xi $ the boundary part of the EMT tends to zero as
$a\to 0$, i.e. $\langle T_i^k\rangle ^{(b)}\to 0$ in the limit
when the barrier coincides with the Rindler horizon. As a result
the corresponding VEV's of the EMT are the same as for the
Fulling-Rindler vacuum without boundaries. Hence, the barrier
located at the Rindler horizon does not alter the vacuum EMT.

The vacuum stresses in the region $0<\xi <a$ are investigated in
Sec. \ref{sec:left} for scalar and electromagnetic fields. The
corresponding boundary parts can be presented as Eqs.
(\ref{emtboundN1}) and (\ref{emtboundel}), respectively. These
formulas differ from the ones for the region $\xi >a$ by the
replacements $I_\omega \to K_\omega , K_\omega \to I_\omega $. For
the scalar field the boundary part diverges at the horizon with
the leading behavior (\ref{asimpxi0}), (\ref{asimpxi0i}). The
divergence of the purely Rindler part is stronger and this part
dominates near the Rindler horizon. Unlike to the scalar case, the
boundary parts of the vacuum EMT components for the
electromagnetic field vanish at the horizon (see asymptotic
formulas (\ref{asimpel0})). In Fig. \ref{fig2el} we present the
vacuum densities for the electromagnetic field, generated by a
perfectly conducting plate in both RR and RL regions. As seen,
unlike to the RR region, where the energy density is negative for
all $a<\xi <\infty $, in the RL region the energy density is
negative near the Rindler horizon, but is positive near the plate.
As a result for some intermediate value of $\xi $ we have zero
energy density.

\section*{Acknowledgements}

I am grateful to Professor E. Chubaryan and Professor A. Mkrtchyan
for general encouredgement and suggestions, and to R. Avagyan, L.
Grigoryan and A. Yeranyan for useful discussions. This work was
supported by the Armenian National Science and Education Fund
(ANSEF) Grant No. PS14-00.

\appendix

\section{Summation formula over zeros of $\bar{K}_{\omega }$}
\label{sec:app1}

In Sec. \ref{sec:vev} we have shown that the VEV's for the EMT
components in the RR region contain sums over zeros $\omega
=\omega _{n}(\eta )$, $n=1,2,\ldots $ of the function
$\bar{K}_{i\omega }(\eta )=AK_{i\omega }(\eta )+bzK_{i\omega
}^{\prime }(\eta )$ for a given $\eta $. To obtain a summation
formula over these zeros we will use the generalized Abel-Plana
formula derived in \cite{Sahmat},\cite{Sahreview}. In this
formula as functions $f(z)$ and $%
g(z)$ let us choose
\begin{eqnarray}
f(z) &=&\frac{2i}{\pi }\sinh \pi z\,F(z),  \label{fgAPF} \\
g(z) &=&\frac{\bar{I}_{iz}(\eta )+\bar{I}_{-iz}(\eta )}{\bar{K}%
_{iz}(\eta )}F(z),  \nonumber
\end{eqnarray}
with a function $F(z)$ analytic in the right half-plane ${\rm
Re}\, z\ge 0$, and quantities with overbars are defined in
accordance with Eq. (\ref{barnot}). For the sum and difference of
these functions one has
\begin{equation}
g(z)\pm f(z)=2F(z)\frac{\bar{I}_{\mp iz}(\eta )}{\bar{K}_{iz}(\eta
)} . \label{fgsum}
\end{equation}
By using the asymptotic formulas for the Bessel modified functions
for large values of index, the conditions for the generalized
Abel-Plana formula can be written in terms of the function $F(z)$
as follows
\begin{equation}
\vert F(z)\vert <\epsilon (\vert z\vert)e^{-\pi x}\left(
\frac{\vert z \vert }{\eta }\right) ^{2\vert y\vert },\quad
z=x+iy,\quad x>0, \quad \vert z\vert \to \infty , \label{condf1}
\end{equation}
where $\vert z\vert \epsilon (\vert z\vert )\to 0$ when $\vert
z\vert \to \infty $. Let $\omega =\omega _n$ be zeros for the
function $\bar K_{i\omega }(\eta )$ in the right half-plane. Here
we will assume values of $A$ and $b$ for which all these zeros are
real. It can be seen that they are simple. This directly follows
from the relation
\begin{equation}\label{relap1simp}
  \frac{\partial }{\partial \omega }\bar{K}_{i\omega}(\eta )\left|
  _{\omega =\omega _n}=\frac{2\omega _n b}{K_{i\omega _n}(\eta )}
  \int_{1}^{\infty }\frac{d\xi }{\xi }K_{i\omega _n}^{2}(\xi \eta
  )\right.,
\end{equation}
which is a direct consequence of integral formula
(\ref{intformgen}) with $\phi _{\omega }^{(1)}(\xi)=\phi _{\omega
}^{(2)}(\xi)=K_{i\omega _n}(\xi \eta )$. Substituting functions
(\ref{fgAPF}) into generalized Abel-Plana formula (formula (2.11)
in Ref. \cite{Sahreview} ) and
 taking into account that the points $z =\omega _n$ are simple
 poles for the function $g(z)$ and the relation $\bar I_{i\omega _n}(\eta
)=\bar I_{-i\omega _n}(\eta )$, one obtains
\begin{eqnarray}
&&\lim_{l\to \infty }\left\{ \sum_{n=1}^{n_l
}\frac{\bar{I}_{i\omega _{n}}
(\eta )F(\omega _{n})}{\partial \bar{%
K}_{i\omega }(\eta )/\partial \omega \vert _{\omega =\omega _{n}}}
-\frac{1}{\pi ^{2}}\int_{0}^{l}\sinh \pi z\,F(z)dz\right\}  \nonumber \\
&& \qquad =
-\frac{1}{2\pi }\int_{0}^{\infty }\frac{\bar{I}_{z}(\eta )}{\bar{K}%
_{z}(\eta )}\left[ F(ze^{\pi i/2})+F(ze^{-\pi i/2})\right] dz ,
\label{sumformula}
\end{eqnarray}
where $\omega _{n_{l}}<l<\omega _{n_{l}+1}$. Here we have assumed
that the function $F(z)$ is analytic in the right half-plane.
However, this formula can be easily generalized for the functions
having poles in this region.

\section{$D=1$ massless case: Direct evaluation} \label{sec:app2}

Here we show that the direct evaluation of the boundary VEV's in
$D=1$ gives the same result as the analytical continuation from
the higher values of the space dimension. Now in equation
(\ref{eqxi}) for the function $\phi (\xi )$ one has $\lambda =0$.
The corresponding linearly independent solutions are $e^{\pm
i\omega \ln (\xi /a)}$. The normalized eigenfunctions satisfying
boundary condition (\ref{boundRind}) are in form
\begin{equation}
\varphi _{\omega }(\xi ,\tau )=\frac{e^{-i\omega \tau }}{\sqrt{\pi \omega }}%
\sin \left[ \omega \ln (\xi /a)-\beta \right] ,\quad 0<\omega <\infty ,
\label{eigD1m0}
\end{equation}
where
\begin{equation}
e^{i\beta }=\frac{A+i\omega B/a}{\sqrt{A^{2}+(\omega B/a)^{2}}}.
\label{D1m0bet}
\end{equation}
Note that these eigenfunctions have the same form for the regions
$\xi >a$ and $\xi <a$. Unlike to $D>1$ cases now the both types of
the eigenfunctions are bounded and the spectrum for $\omega $ in
the region $\xi
>a$ is continous. Substituting eigenfunctions (\ref{eigD1m0}) into
the mode sum formula (\ref{Wightvev}) after some algebra one finds
the Wightman function in the form (\ref{WRb}), where the first
summand on the right is the $D=1$ Wightman function for the
Rindler wedge without boundaries:
\begin{equation}
G_R^+(x,x')=\frac{1}{2\pi }\int _{0}^{\infty }\frac{d\omega }{\omega }
e^{-i\omega (\tau -\tau ')}\cos
{[ \omega \ln ( \xi /\xi ')] },
\label{WRD1}
\end{equation}
and for the boundary part one has
\begin{equation}
\langle \varphi (x)\varphi (x')\rangle ^{(b)}=-\frac{1}{2\pi }
\int _{0}^{\infty }\frac{d\omega }{\omega }
e^{-i\omega (\tau -\tau ')}\cos
{[ \omega \ln (\xi \xi '/a^2) -2\beta ] }.
\label{WbD1}
\end{equation}
To transform this expression we write the cos function in terms of
exponentials and rotate the integration contour by angle $\pm \pi
/2$ for the term with ${\rm exp}[i\omega \ln (\xi \xi '/a^2)]$ and
by angle $\mp \pi/2$ for the term with ${\rm exp}[-i\omega \ln
(\xi \xi '/a^2)]$. Here and below the upper and lower signs
correspond to the cases $\xi, \xi '>a$ and $\xi ,\xi '<a$
respectively. Under the condition $|\tau -\tau '|<|\ln (\xi \xi
'/a^2)|$ the integrals over the arcs with large radius in the
complex $\omega $-plane tend to zero and we receive
\begin{equation}
\langle \varphi (x)\varphi (x')\rangle ^{(b)}=-\frac{1}{2\pi }
\int _{0}^{\infty }\frac{d\omega }{\omega }\frac{A\pm \omega
B/a}{A \mp \omega B/a}e^{-\omega |\ln (\xi \xi '/a^2)|}\cosh
[\omega (\tau - \tau ')] .\label{WbD11}
\end{equation}
The boundary contribution to the vacuum EMT can be found
substituting this expression into formula (\ref{vevemtW}). It can
be easily seen that as a result for the boundary part we obtain
formulas (\ref{D1boundm0}) and (\ref{D1m0left}). Hence, we have
shown that the direct evaluation gives the same result as the
analytic continuation.

\section{Proof of identity {\protect{(\ref{IDENTITY2})}}} \label{sec:app3}

In this appendix we will prove identity (\ref{IDENTITY2}), which
was used to see the equivalency of our formulas
(\ref{epsWD1})-(\ref{piWD}) for the VEV's of the EMT in the
special cases of the conformally and minimally coupled massive
scalars to the results derived by Hill in Ref. \cite{Hill}. Our
starting point is the integral
\begin{equation}\label{intL}
  {\mathcal{L}}(\xi ,\xi ')=\int_{0}^{\infty }dk \, k^{D-2}\int_{0}^{\infty
  }d\omega \, \omega ^{2}e^{-\pi \omega }K_{i\omega }(\lambda \xi
  )K_{i\omega}(\lambda \xi ')
\end{equation}
with $\lambda $ defined as in Eq. (\ref{eqxi}). By using formulas
(\ref{intK2}) and \cite{Prudnikov2}
\begin{equation}\label{intap3}
\int_{0}^{\infty }dk \, k^{D-2}K_{0}\left( \gamma \sqrt{k^2+m^2}
\right)=2^{(D-3)/2}m^{D-1}\Gamma \left( \frac{D-1}{2} \right)
\frac{K_{(D-1)/2}(m\gamma )}{(m\gamma )^{(D-1)/2}},
\end{equation}
for this integral one obtains
\begin{equation}\label{form1L}
{\mathcal{L}}(\xi ,\xi ')=2^{(D-1)/2}\pi m^{D-1}\Gamma \left(
\frac{D-1}{2} \right) \int_{0}^{\infty }dy\, \frac{\pi
^2-3y^2}{(\pi ^2+y^2)^3}\frac{K_{(D-1)/2}(m\gamma )}{(m\gamma
)^{(D-1)/2}},
\end{equation}
where $\gamma $ is defined by Eq. (\ref{intK2}). In particular
taking $\xi '=\xi $ we receive
\begin{equation}\label{form1Leq}
{\mathcal{L}}(\xi ,\xi )=2^{(D-1)/2}\pi m^{D-1}\Gamma \left(
\frac{D-1}{2} \right) \int_{0}^{\infty }dy\, \frac{\pi
^2-3y^2}{(\pi ^2+y^2)^3}\frac{K_{(D-1)/2}(z)}{z^{(D-1)/2}},
\end{equation}
with $z$ defined in accordance to Eq.(\ref{phi2dif2}). Another
form for $\mathcal{L}(\xi ,\xi ')$ can be obtained by using the
relation
\begin{equation}\label{relap13}
  \omega ^{2}K_{i\omega }(\lambda \xi )=\lambda ^{2}\xi ^{2}K_{i\omega }(\lambda \xi )
  -\xi ^{2}\frac{d^{2}K_{i\omega }(\lambda \xi )}{d\xi ^{2}}-
  \xi \frac{dK_{i\omega }(\lambda \xi )}{d\xi }.
\end{equation}
The substitution into Eq. (\ref{intL}) yields
\begin{eqnarray}\label{form2L}
{\mathcal{L}}(\xi ,\xi ')&=& 2^{(D-3)/2}\pi m^{D+1}\Gamma \left(
\frac{D-1}{2} \right) \int_{0}^{\infty }\frac{dy}{\pi ^2+y^2}
\left\{ (D-1)\frac{K_{(D+1)/2}(m\gamma )}{(m\gamma
)^{(D+1)/2}}\right. \\
& & \left. +\left[ 1-\frac{d^2}{d(m\xi )^2}-\frac{1}{m\xi
}\frac{d}{d(m\xi )}\right]\frac{K_{(D-1)/2}(m\gamma )}{(m\gamma
)^{(D-1)/2}}\right\} . \nonumber
\end{eqnarray}
For the coincidence case $\xi '=\xi $ from here one obtains
\begin{eqnarray}\label{form2Leq}
{\mathcal{L}}(\xi ,\xi )&=& 2^{(D-5)/2}\pi m^{D+1}\xi ^2\Gamma
\left( \frac{D-1}{2} \right) \int_{0}^{\infty }\frac{dy}{\pi
^2+y^2} \left\{ \left[ D+1-(D-1)\cosh y\right]
\frac{K_{(D+1)/2}(z)}{z^{(D-1)/2}}\right. \\
& & \left. +(1-\cosh y)\frac{K_{(D-1)/2}(z)}{z^{(D-1)/2}}\right\}.
\nonumber
\end{eqnarray}
Comparing expressions (\ref{form1Leq}) and (\ref{form2Leq}) one
obtains identity (\ref{IDENTITY2}).

\end{document}